%% file: 38335.tex
\documentclass{aa}  

\usepackage{pgffor}
\usepackage{amssymb,times,graphicx, subfigure}
\usepackage[english]{babel}
\usepackage[varg]{txfonts}
\usepackage{xspace}
\usepackage{longtable,lscape,supertabular}
\usepackage{rotating}
\usepackage{color}
\usepackage{etoolbox}

\usepackage{natbib,afterpage,twoopt}
\bibpunct{(}{)}{;}{a}{}{,} 

\def\irc{IRC\,+10\,216\xspace}
\def\iktau{IK~Tau\xspace}
\def\vycma{VY~CMa\xspace}
\def\cit6{CIT~6\xspace}

\def\whya{W~Hya\xspace}
\def\waql{W~Aql\xspace}
\def\vaql{V~Aql\xspace}

\def\txcam{TX~Cam\xspace}
\def\rdor{R~Dor\xspace}
\def\nmlcyg{NML~Cyg\xspace}
\def\chicyg{$\chi$~Cyg\xspace}
\def\pigru{$\pi^1$~Gru}
\def\um{$\mu$m\xspace}

\def\rstar{$R_{\star}$\xspace}
\def\msun{$M_{\sun}$\xspace}

\def\msunyr{$M_{\sun}$\,yr$^{-1}$\xspace}
\def\kms{km\,s$^{-1}$\xspace}

\def\mdot{$\dot{M}$\xspace}
\def\twth{$^{12}$C/$^{13}$C\xspace}
\def\twco{$^{12}$CO\xspace}
\def\thco{$^{13}$CO\xspace}

\usepackage{array}
\newcolumntype{L}[1]{>{\raggedright\let\newline\\\arraybackslash\hspace{0pt}}m{#1}}
\newcolumntype{C}[1]{>{\centering\let\newline\\\arraybackslash\hspace{0pt}}m{#1}}
\newcolumntype{R}[1]{>{\raggedleft\let\newline\\\arraybackslash\hspace{0pt}}m{#1}}

\begin{document}

   \title{The surprisingly carbon-rich environment of the S-type star \waql \thanks{This publication is based on data acquired with the Atacama Pathfinder Experiment (APEX). APEX is a collaboration between the Max-Planck-Institut f\"{u}r Radioastronomie, the European Southern Observatory, and the Onsala Space Observatory.}}

   \author{E. De Beck
          \and H. Olofsson
          }

   \institute{Department of Space, Earth and Environment, Chalmers University of Technology, Onsala Space Observatory, 43992 Onsala, Sweden
                  \\ \email{elvire.debeck@chalmers.se}
              }

   \date{Received 4 May 2020 / Accepted 3 July 2020}

  \abstract
   {
\waql is an asymptotic giant branch (AGB) star with an atmospheric elemental abundance ratio $\mbox{C/O}\approx0.98$. It has previously been reported to have circumstellar molecular abundances intermediate between those of M-type and C-type AGB stars, which respectively have $\mbox{C/O}<1$ and $\mbox{C/O}>1$. This intermediate status is considered typical for S-type stars, although our understanding of the chemical content of their circumstellar envelopes is currently rather limited.
   }
   {
We wish to assess the reported intermediate status of \waql by analysing the line emission of molecules that have not been observed towards this star before.
   }
   {
We have performed observations in the frequency range $159-268$\,GHz with the SEPIA/B5 and PI230 instruments on the APEX telescope. We make abundance estimates through direct comparison to available spectra towards a number of well-studied AGB stars and based on rotational diagram analysis in the case of one molecule.
   }
   {
From a compilation of our abundance estimates and those found in the literature for two M-type (\rdor, \iktau), two S-type (\chicyg, \waql), and two C-type stars (\vaql, \irc), we conclude that \waql's circumstellar environment appears considerably closer to that of a C-type AGB star than to that of an M-type AGB star. In particular, we detect emission from C$_2$H, SiC$_2$, SiN, and HC$_3$N, molecules previously only detected towards the circumstellar environment of C-type stars. This conclusion, based on the chemistry of the gaseous component of the circumstellar environment, is further supported by reports in the literature on the presence of atmospheric molecular bands and spectral features of dust species which are typical for C-type AGB stars. Although our observations mainly trace species in the outer regions of the circumstellar environment, our conclusion matches closely that based on recent chemical equilibrium models for the inner wind of S-type stars: the atmospheric and circumstellar chemistry of S-type stars likely resembles that of C-type AGB stars much more closely than that of M-type AGB stars. 
   }
  {
Further observational investigation of the gaseous circumstellar chemistry of S-type stars is required to characterise its dependence on the atmospheric $\mbox{C/O}$.  Non-equilibrium chemical models of the circumstellar environment of AGB stars need to address the particular class of S-type stars and the chemical variety that is induced by the range in atmospheric $\mbox{C/O}$. 
  }
  

   \maketitle



\section{Introduction}\label{sect:introduction}
Evolved stars on the asymptotic giant branch (AGB) are typically classified according to the abundance ratio of carbon and oxygen atoms, C/O, in combination with a set of signatures pertaining to molecular bands. Strong molecular bands of titanium monoxide, TiO, are seen in the atmospheres of M-type AGB stars ($\mathrm{C/O}<1$); zirconium oxide, ZrO, bands are seen for S-type stars ($\mathrm{C/O}\approx1$); and bands of carbon compounds, such as CN, are indicative of C-type ($\mathrm{C/O}>1$) stars \citep{agbbook}.

The atmospheric content of S-type stars, and in particular the abundance ratio TiO/ZrO, is critically linked to the value of the C/O abundance ratio \citep[e.g.][]{smolders2012_S_SpitzerSurvey} and one can expect a similar link to exist for the contents of the circumstellar envelopes (CSEs) of these stars, built up by the stellar mass loss. \citet{hony2009_S_ISO} and \citet{smolders2010_pah} showed that the dust around S-type stars can show a mixture of dust species typical for M-type and C-type stars, including silicates, MgS, and possibly even polycyclic aromatic hydrocarbons (PAHs). 

Chemical models of the gas in AGB CSEs reported in the literature focus primarily on M-type and C-type stars \citep[e.g.][]{agundez2006_Ochemistry_irc10216,agundez2012_innerlayers_irc10216, cherchneff2006,cherchneff2012, cordiner2009, gobrecht2016_dustformation_iktau, vandesande2018, willacy1997}. Recently, \cite{agundez2020_chemequilibrium} presented chemical equilibrium models for S-type stars, but non-equilibrium chemical models for S-type stars with a similar coverage of molecular species are currently lacking.  Additionally, the available chemical models assume high mass-loss rates ($\approx10^{-5}$\,\msunyr; most often using the M-type \iktau and the C-type \irc as reference stars) and leave a clear gap in the description of the chemical content of CSEs of low- to intermediate-mass-loss rate stars ($10^{-8}-10^{-6}$\,\msunyr). 

Sample studies of the molecular content of the CSEs of S-type stars are currently limited to targeted observations of standard molecules, such as CO, SiO, SiS, HCN, and CS \citep{danilovich2018_cs_sis,ramstedt2014_12co13co,ramstedt2009_sio,schoeier2013_hcn}. Sample sizes vary from six to forty across these studies. The gas chemical content of the CSEs of the S-type AGB stars \waql and \chicyg has been studied in more detail by \citet{decin2008_parentmolecules}, \citet{schoeier2011_chicyg}, \citet{danilovich2014}, and \citet{brunner2018_waql_alma} using ALMA and Herschel observations, adding H$_2$O, NH$_3$, and CN to the list of molecules detected towards S-type AGB stars. Unbiased observations of the CSEs of S-type stars, not targeting particular molecules but covering large bandwidths to also detect ``unexpected'' species, have not yet been presented in the literature.

We present observations at long wavelengths ($1-2$\,mm) of molecular line emission from the CSE of the S-type AGB star \object{W~Aql}. This star has a reported $\mathrm{C/O}\approx0.98$ \citep{keenan1980_Stype,danilovich2015_waql_binary} and shows a mix of M-type and C-type signatures in its dust spectrum \citep{hony2009_S_ISO}. \waql is a Mira-type variable with a pulsation period of 490 days \citep{alfonsogarzon2012}, is located at an estimated distance of 395\,pc \citep[][and references therein]{danilovich2014}, and has a mass-loss rate of $3\times10^{-6}$\,\msunyr \citep{ramstedt2017_waql}. 

We discuss the molecular footprint of \waql's CSE, how this compares to what we know about its atmosphere and dust, and put this in relation to what is known about CSEs of ``prototypical'' M-type (\rdor and \iktau) and C-type (\irc, \vaql) stars.

\section{Observations}\label{sect:observations}

\begin{figure*}
\includegraphics[width=\linewidth]{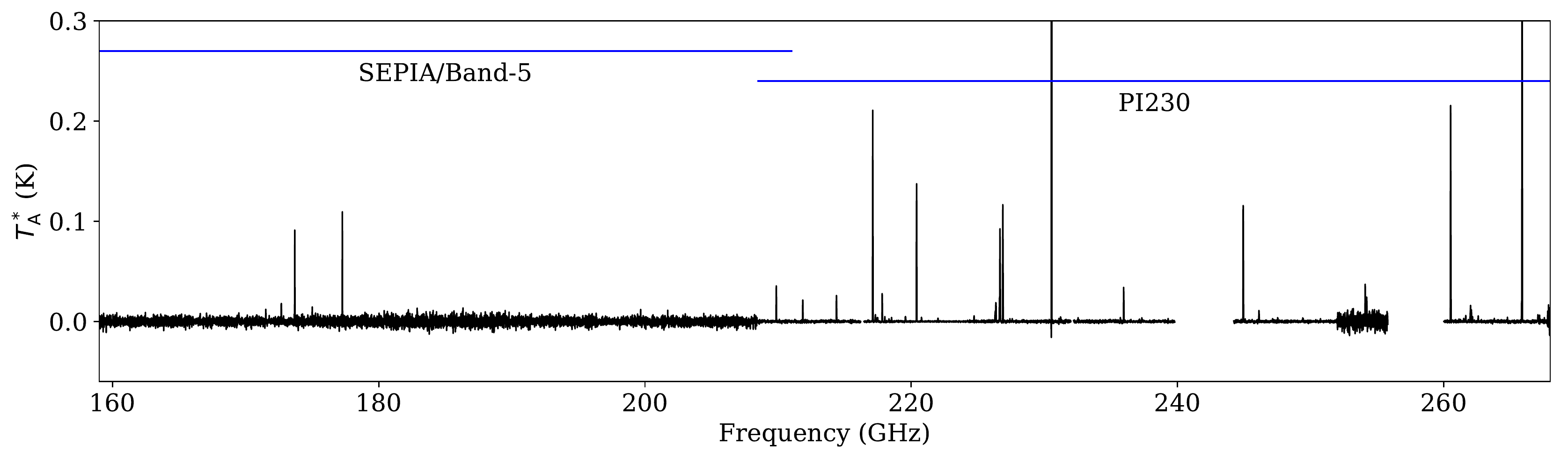} 
\caption{APEX observations towards \waql. The frequency ranges covered by the SEPIA/B5 and PI230 instruments are marked with blue horizontal lines. \label{fig:overviewfig}}
\end{figure*}

In the framework of a broader project in which the effect of AGB wind density on the circumstellar chemistry is studied, we carried out a spectral survey of \waql in the range $159-211$\,GHz using Band 5 of the Swedish-ESO PI Instrument \citep[SEPIA;][]{billade2012_band5,belitsky2018_sepia} on the Atacama Pathfinder Experiment telescope \citep[APEX; ][]{gusten2006_apex}. The observations were carried out on 9 May 2017 in very good weather, at $0.8-1.5$\,mm of precipitable water vapour (PWV). We obtained follow-up observations using the APEX/PI230 instrument\footnote{PI230 is a collaboration between the European Southern Observatory (ESO) and the Max-Planck-Institut für Radioastronomie (MPIfR).} on 17 and 21 May, 21 and 22 August, and 21 October 2018 with a PWV in the range $0.7-1.0$\,mm, and on 2, 3, 7, and 11 July 2019 with a PWV in the range $0.9-5.8$\,mm. 

All observations were carried out in wobbler-switching mode using the standard 50\arcsec\/ beam throw. The size of the main beam is $30-39$\arcsec\/ at $159-211$\,GHz (SEPIA/B5) and $27-30$\arcsec\/ at $209-236$\,GHz and $23-25$\arcsec\/ at $244-272$\,GHz (PI230). 

We process the fully calibrated data with the \textsc{gildas/class}\footnote{\url{http://www.iram.fr/IRAMFR/GILDAS/}} package. After removing bad channels, which are typically present at $<50$\,MHz from the band edges, and masking relevant spectral features we remove a first-degree polynomial baseline from the spectra before averaging. We show the resulting spectra in $T_{\mathrm{A}}^*$, the antenna temperature corrected for spillover and atmospheric losses, throughout the paper. A snapshot overview is shown in Fig.~\ref{fig:overviewfig}; for detailed spectra we refer to  Figs.~\ref{fig:fullscan_waql_b5} (SEPIA/B5) and \ref{fig:fullscan_waql_combined} (PI230) in the appendix. We refer to App.~\ref{app:data} for a brief discussion on the occurrence of a small number of features in the spectra that originate from sideband contamination.

Noise characteristics vary significantly throughout the covered spectral range and are summarised in Table~\ref{tbl:rms}. These values are approximate, as the rms noise can vary somewhat depending on the exact location in the spectrum as a consequence of different observing conditions (see above). The increase in rms noise in the SEPIA//B5 range is caused by the interference of atmospheric water around 183\,GHz. The high noise in the range $252.0-255.8$\,GHz is a consequence of the short integration time at this frequency, connected to tuning problems. We include the data as they include several detected lines. The frequency resolution at which the rms noise values in Table~\ref{tbl:rms} are measured translates to a velocity resolution in the range $5.5-6.5$\,\kms, allowing for detection of circumstellar spectral lines towards \waql (see Sect.~\ref{sect:results}).

We have adopted a systemic velocity $\varv_{\rm sys}=-22.8$\,\kms for \waql and all velocities $\varv$ reported here are relative to this value within the Local Standard of Rest (LSR) framework, that is $\varv = \varv_{\rm LSR} - \varv_{\rm sys}$.

\begin{table}
\caption{Noise characteristics. We list approximate rms noise across different frequency intervals in our data for frequency resolution $\Delta\nu$. \label{tbl:rms}}
\centering
\begin{tabular}{lccc}
\hline\hline\\[-2ex]
Instrument	&Frequency range& Rms 	& $\Delta\nu$ \\
&(GHz) & (mK)	& (MHz)\\
\hline\\[-2ex]
SEPIA/B5 &$159.0 - 178.0$				& 3.5				&3.2\\
&$178.0 - 188.0$				& 5.0				&3.2\\
&$188.0 - 208.4$				& 3.5				&3.2\\
PI230&$208.4 - 252.0$				& 0.5				&4.3\\
&$252.0 - 255.8$ & 5.5 & 5.1\\
&$260.0 - 267.8$				& 0.6				&5.4\\
\hline\hline
\end{tabular}
\end{table}

\begin{table}
\begin{center}
\caption{Molecules identified in our survey towards \waql. Line numbers listed in column 3 correspond to those in Column 1 of Table~\ref{tbl:fullscan_waql_combined}. We mark both new and tentative detections for clarity. \label{tbl:mol_summary}}
\begin{tabular}{lC{1.5cm}L{4.25cm}}
\hline \hline \\[-2ex]
Molecule & Number of lines & Line Numbers \\
\hline \\[-2ex]
$^{13}$CN \tablefootmark{$\ddagger$} & 3 & 24, 25, 26 \\
$^{13}$CO & 1 & 33 \\
$^{13}$CS \tablefootmark{$\ddagger$} & 1 & 43 \\
$^{29}$SiO & 2 & 4, 22 \\
$^{29}$SiS \tablefootmark{$\ddagger$} & 3 & 21, 60, 83 \\
$^{30}$SiO \tablefootmark{$\ddagger$} & 2 & 17, 63 \\
$^{30}$SiO \tablefootmark{$\dagger$} \tablefootmark{$\ddagger$} & 1 & 3 \\
$^{30}$SiS & 3 & 16, 41, 76 \\
$^{30}$SiS \tablefootmark{$\dagger$} & 1 & 53 \\
C$^{17}$O & 1 & 36 \\
C$^{18}$O & 1 & 31 \\
C$^{34}$S \tablefootmark{$\dagger$} \tablefootmark{$\ddagger$} & 1 & 12 \\
C$_2$H \tablefootmark{$\ddagger$} & 2 & 72, 73 \\
C$_2$H \tablefootmark{$\dagger$} \tablefootmark{$\ddagger$} & 3 & 8, 9, 75 \\
CN & 3 & 37, 38, 39 \\
CO & 1 & 42 \\
CS & 1 & 52 \\
H$^{13}$CN & 1 & 5 \\
HC$_3$N \tablefootmark{$\dagger$} \tablefootmark{$\ddagger$} & 7 & 6, 14, 29, 40, 48, 54, 77 \\
HCN & 2 & 10, 80 \\
HCN, $v_2=1$ & 2 & 79, 82 \\
HCO$^+$ \tablefootmark{$\dagger$} \tablefootmark{$\ddagger$} & 2 & 11, 84 \\
HN$^{13}$C \tablefootmark{$\ddagger$} & 1 & 68 \\
NaCl \tablefootmark{$\dagger$} \tablefootmark{$\ddagger$} & 2 & 57, 65 \\
Si$^{17}$O \tablefootmark{$\ddagger$} & 1 & 61 \\
Si$^{34}$S \tablefootmark{$\ddagger$} & 3 & 18, 56, 78 \\
SiC$_2$ \tablefootmark{$\ddagger$} & 13 & 19, 20, 34, 35, 44, 45, 46, 49, 50, 58, 67, 69, 71 \\
SiC$_2$ \tablefootmark{$\dagger$} \tablefootmark{$\ddagger$} & 1 & 64 \\
SiN \tablefootmark{$\ddagger$} & 4 & 28, 30, 70, 74 \\
SiO & 3 & 7, 23, 66 \\
SiS & 4 & 13, 27, 47, 62 \\
SiS \tablefootmark{$\dagger$} & 1 & 2 \\
u & 2 & 32, 59 \\
\hline
\end{tabular}
\tablefoot{\tablefoottext{$\dagger$}{Tentative detection or identification.} \tablefoottext{$\ddagger$}{Species not previously reported towards \waql.}}\end{center}
\end{table}

\begin{figure*}
\centering
\includegraphics[width=0.8\linewidth]{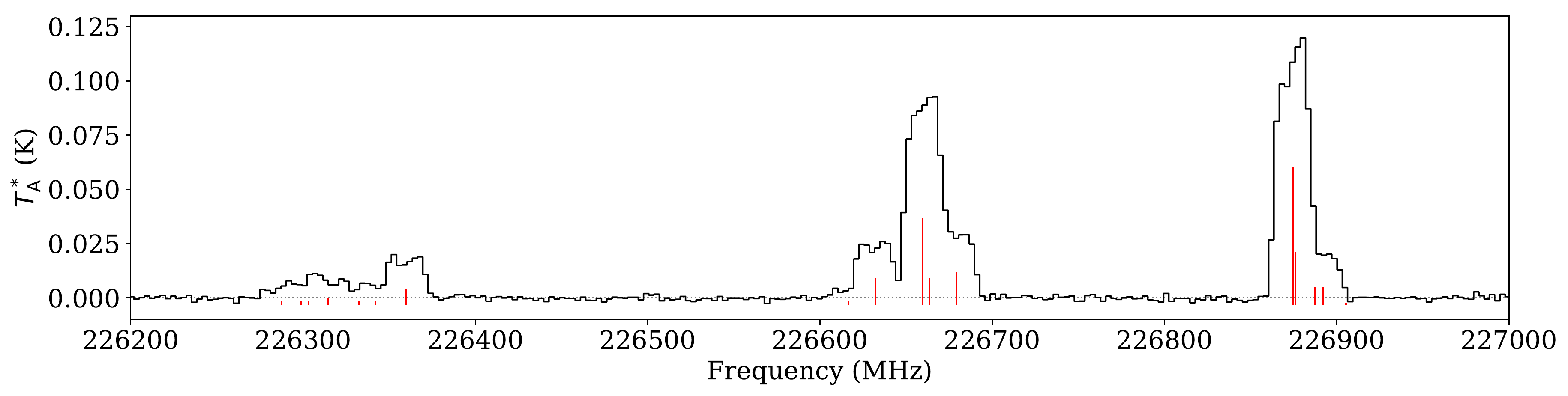}\\ 
\includegraphics[width=0.8\linewidth]{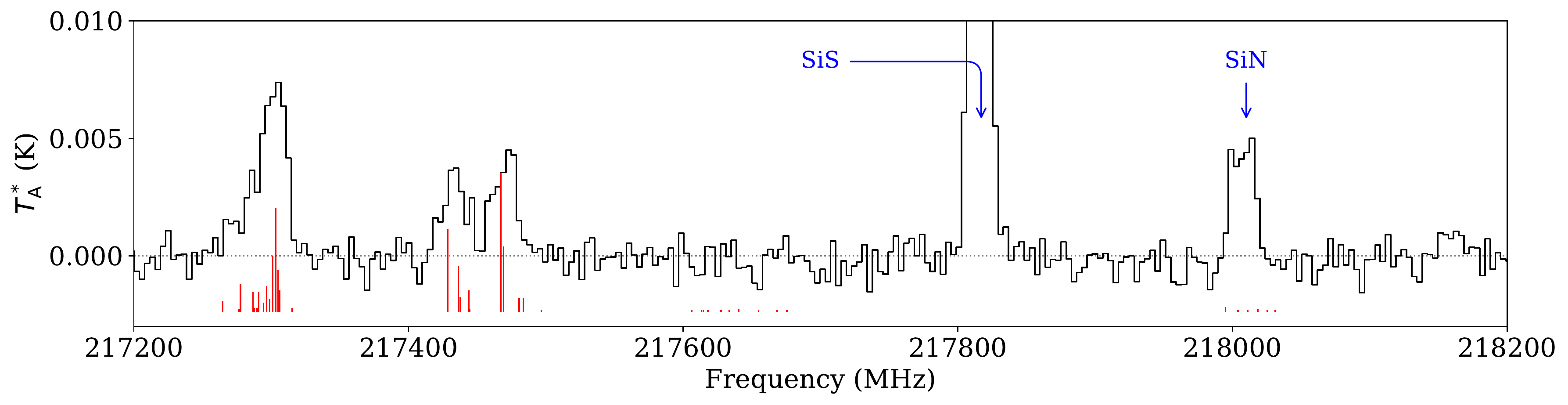} 
\caption{CN emission; \textit{top:} $^{12}$CN, \textit{bottom:} $^{13}$CN. Vertical red lines indicate the position and relative strength of the hyperfine structure components of the rotational transition $2-1$.\label{fig:CN}}
\end{figure*}

\section{Results}\label{sect:results}
We show the entire data set in Figures~\ref{fig:fullscan_waql_b5} (SEPIA/B5) and \ref{fig:fullscan_waql_combined} (PI230) and provide an overview of all measured emission features and their main characteristics in Table~\ref{tbl:fullscan_waql_combined}. We summarise the identified molecules in Table~\ref{tbl:mol_summary}. 

We retain spectral features if the signal in at least three out of six consecutive channels of width $\Delta\nu$ exceeds $3\sigma$ with $\sigma$ and $\Delta\nu$ equal to the rms and frequency resolution listed in Table~\ref{tbl:rms}. This allows us to recover spectral lines over a total velocity interval of roughly two times the terminal velocity of \waql, $v_{\infty}\approx16.5$\,\kms \citep{danilovich2014}. 

All line identifications stem from the Cologne Database for Molecular Spectroscopy\footnote{\url{https://www.astro.uni-koeln.de/cdms/}} \citep[CDMS;][]{mueller2001_cdms,mueller2005_cdms}. We have also used the catalogue for molecular line spectroscopy hosted by the Jet Propulsion Laboratory\footnote{\url{http://spec.jpl.nasa.gov/}} \citep[JPL;][]{pickett1998_jpl} and Splatalogue\footnote{\url{http://www.cv.nrao.edu/php/splat/advanced.php}} as additional resources. Some features are reported as tentative detections although they may not strictly fulfill the above criterion. This is the case if we expect their presence based on other transitions or hyperfine components of the same transition of a given molecule and we find a plausible signal at a lower signal-to-noise ratio. Moreover, the combination of tentative detections of multiple transitions of a given molecule can lead to a firm detection of that molecule, as we discuss in Sect.~\ref{sect:newdetections} in the case of HC$_3$N.

In Sect.~\ref{sect:inventory} we summarise which molecules we identify in our observations, how our results compare to earlier reports on CSEs of M-, S-, and C-type AGB stars in general, and on the CSE of \waql in particular. In Sect.~\ref{sect:isotopes} we shortly discuss the detection or non-detection of different isotopologues for some of the main elements.

\subsection{Molecular inventory \label{sect:inventory}}
We detect emission from transitions of CO, SiO, HCN, CS, SiS, and CN as expected for the CSE of an S-type AGB star and consistent with previous reports for \waql  \citep[e.g.,][]{brunner2018_waql_alma,danilovich2014,danilovich2018_cs_sis,schoeier2011_chicyg}. We additionally detect emission from species considered to be typical for, or until now only detected in, the CSEs of carbon-rich AGB stars: SiC$_2$, SiN, C$_2$H, and the cyanopolyyne HC$_3$N. We give an overview of these detections and provide some first abundance estimates to relate our findings to those made for M-type and C-type CSEs. We briefly address some tentative detections and unidentified emission features. 

\subsubsection{Previously detected: CO, SiO, HCN, CS, SiS, CN}
CO, SiO, HCN, and CS are expected to form in relatively high abundances in the inner winds of all chemical types of AGB stars \citep{cherchneff2006,agundez2020_chemequilibrium}. SiS emission is also commonly detected towards CSEs around all chemical types of AGB stars \citep{schoeier2013_hcn}. We refer to the literature for detailed discussions on these molecules in the CSE of \waql: \citet{ramstedt2014_12co13co} and \citet{danilovich2014} present radiative transfer models of CO emission and \citet{brunner2018_waql_alma} present models of SiO, HCN, CS, and SiS.

\paragraph{CO} We detect emission from $^{12}$CO\,($J=2-1$) and $^{13}$CO\,($J=2-1$), see Fig.~\ref{fig:CO}, in agreement with the lines published by \citet{debeck2010_comdot} and \citet{ramstedt2014_12co13co} and with the calibration spectra made available by the APEX observatory\footnote{Spectra can be retrieved from \url{http://www.apex-telescope.org/heterodyne/shfi/calibration/database}}. 

\paragraph{SiO}
We detect three  emission lines from the main SiO isotopologue ($J=4-3,5-4,6-5$) and one from Si$^{17}$O\,($J=6-5$) at 250.7\,GHz. We do not detect emission from Si$^{18}$O, in line with the intensity of the Si$^{17}$O emission and the sensitivity reached in the spectra covering the relevant transitions of Si$^{18}$O assuming $^{17}$O/$^{18}$O\,=\,1.2  \citep{denutte2016_17o18o}. We detect the $^{29}$SiO\,($J=4-3,5-4$) and $^{30}$SiO\,($J=4-3,5-4,6-5$) lines. 
An overview is shown in Fig.~\ref{fig:SiO}.

\paragraph{HCN}
We detect emission from HCN in the vibrational ground state $v=0$ ($J=2-1,3-2$), from both $l$-type doubling components of  $J=3-2$ in the excited bending mode $v_2=1$, and from H$^{13}$CN ($J=2-1$); see Fig.~\ref{fig:HCN}.

\begin{figure*}
\centering
\includegraphics[width=0.96\linewidth]{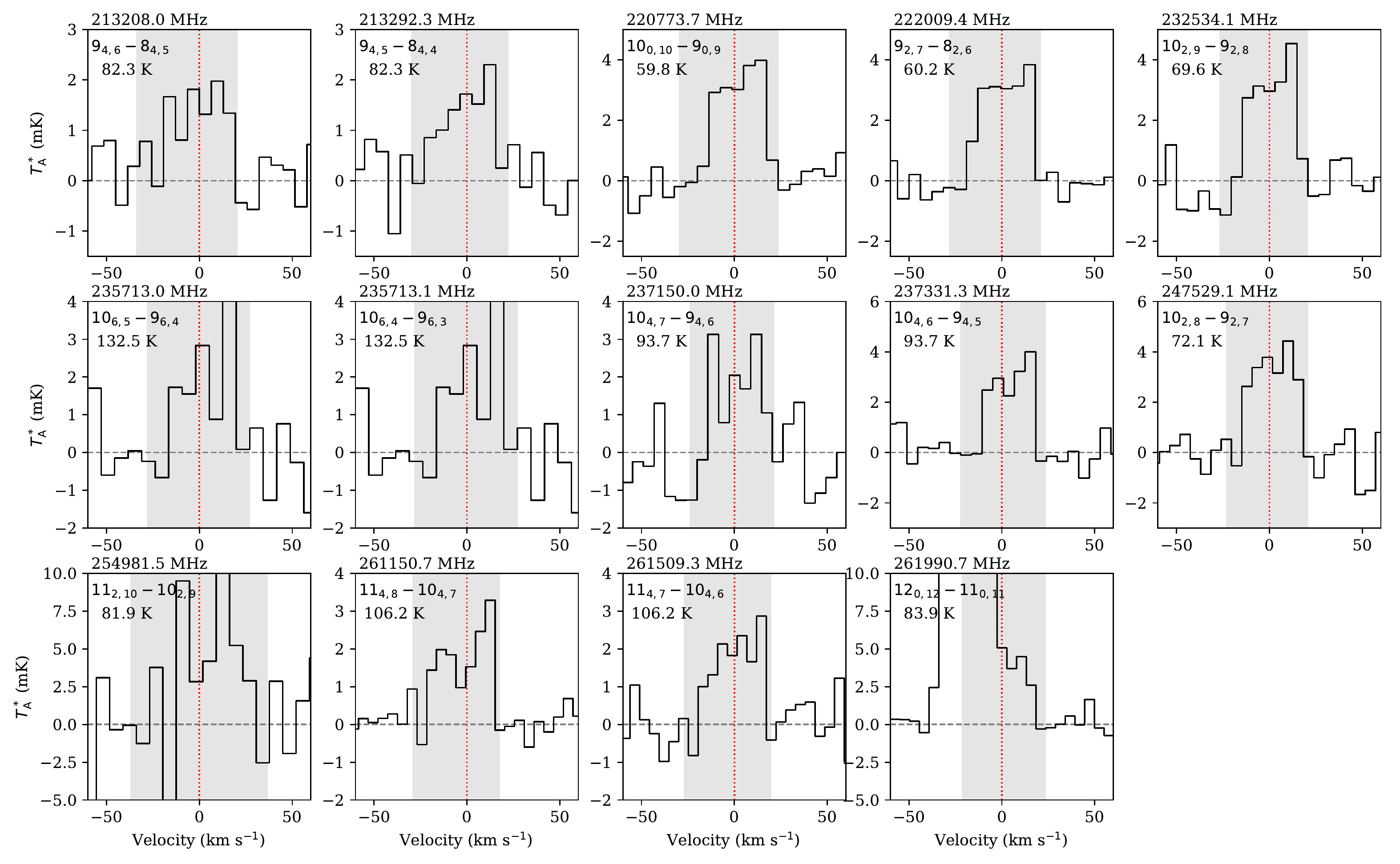} 
\caption{SiC$_2$ emission. Shaded areas indicate the part of the spectrum integrated to obtain the intensities used to construct the rotational diagram in Fig.~\ref{fig:sic2_rotdiagram}, vertical red lines correspond to the rest frequency, centred to a velocity of 0\,\kms. We give the rest frequency, quantum numbers, and upper level energy for each transition. Note: \emph{(i)} the rest frequencies of the $10_{6,5}-9_{6,4}$ and $10_{6,4}-9_{6,3}$ lines at 235.7\,GHz differ by only 0.1\,MHz and \emph{(2)} the emission from the $12_{0,12}-11_{0,11}$ transition at 261.991\,GHz (last panel) is blended with intense C$_2$H emission (Fig.~\ref{fig:cch}) and is therefore omitted from the rotational diagram analysis, as is the $11_{2,10}-10_{2,9}$ transition which is tentatively detected in a region of high rms noise. 
 \label{fig:SiC2}}
\end{figure*}

\begin{figure}
\centering
\includegraphics[width=.8\linewidth]{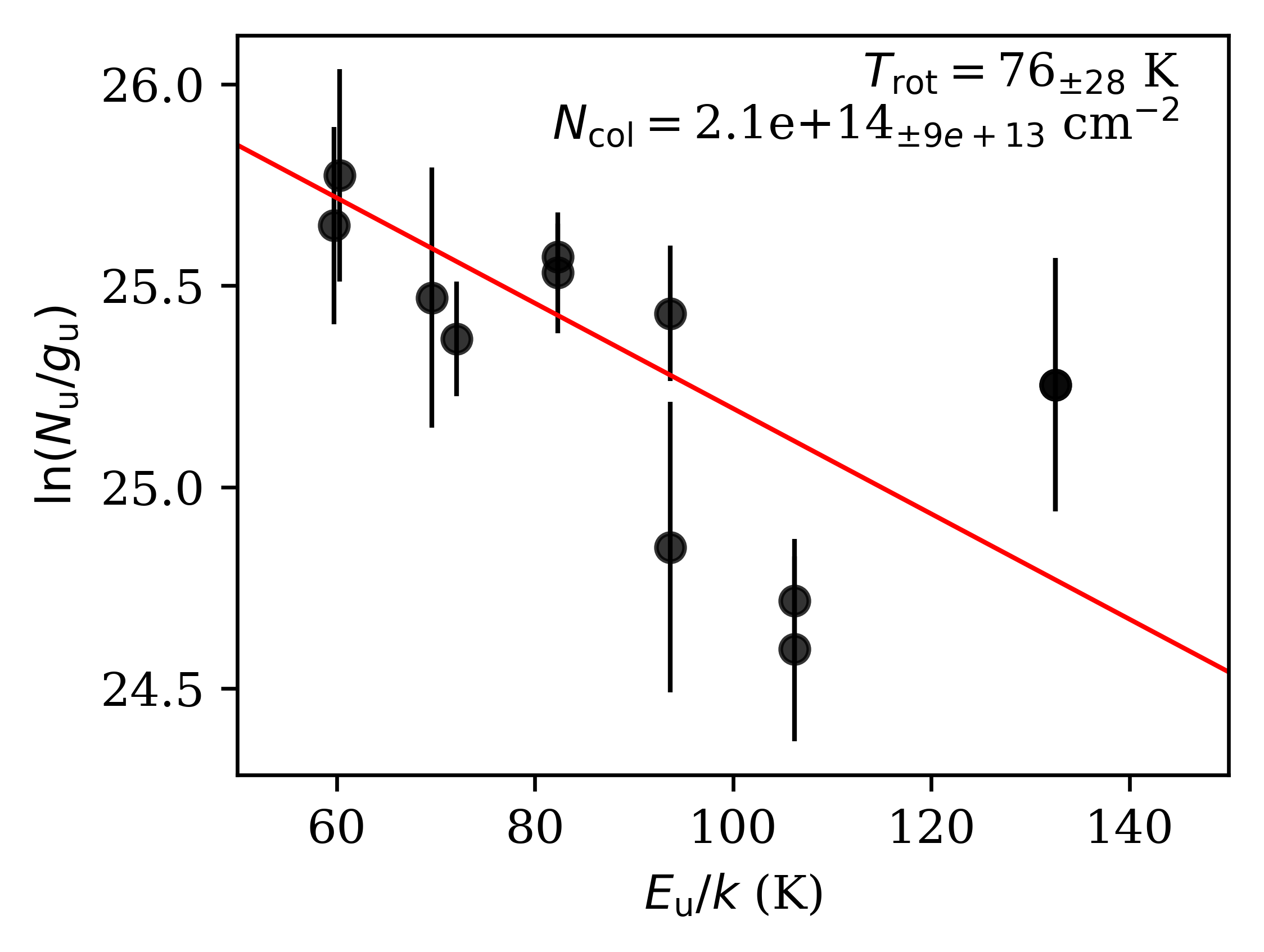} 
\caption{Rotational temperature diagram for SiC$_2$. \label{fig:sic2_rotdiagram}}
\end{figure}

\paragraph{CS}
We detect emission from CS\,($J=5-4$) at 244.9\,GHz with a peak intensity of 115\,mK ($\approx4.5$\,Jy); see Fig.~\ref{fig:CS}. We do not detect emission from CS\,($J=4-3$) at 196.0\,GHz and can set an upper limit to the peak intensity of 12\,mK ($\approx0.5$\,Jy). The $J=5-4$ intensity is roughly in line with the observed intensities of the $J=3-2,6-5,7-6$ lines reported by \citet{brunner2018_waql_alma}, whereas the $J=4-3$ upper limit lies far below the expectations. \citet{cernicharo2014_timevariability} report that low-$J$ CS emission lines show only small variability over time in the case of \irc, so we believe that time variability in the line excitation does not explain the observed discrepancy. Since we do not perform radiative transfer modelling for this paper, we wish to not hypothesise on the reason for the non-detection of the $J=4-3$ line. 

We detect emission from $^{13}$CS\,($J=5-4$) at 231.2\,GHz at a peak intensity of $\approx5$\,mK 
and can set an upper limit of 7\,mK to the peak intensity of $^{13}$CS\,($J=4-3$) at 185.0\,GHz. We also detect emission from C$^{34}$S\,($J=4-3$) at 192.8\,GHz. 

\paragraph{SiS}
We detect emission from the SiS transitions $J=9-8,11-10,12-11,13-12,14-13$ but do not detect SiS\,($J=10-9$) at 181.5\,GHz owing to the higher noise in this region of the SEPIA/B5 spectrum, see Figs.~\ref{fig:fullscan_waql_b5} and \ref{fig:fullscan_waql_combined} and Table~\ref{tbl:rms}. We also detect emission from the isotopologues $^{29}$SiS ($J=12-11, 14-13, 15-14$), $^{30}$SiS ($12-11,13-12, 14-13,15-14$), and Si$^{34}$S ($J=12-11,14-13,15-14$), see Fig.~\ref{fig:SiS}.

\paragraph{CN}
We detect emission over the full hyperfine structure of CN $N=2-1$ between 226.2\,GHz and 227.0\,GHz and of $^{13}$CN $N=2-1$ between 217.2\,GHz and 217.5\,GHz (Fig.~\ref{fig:CN}). Whereas we report the first detection of $^{13}$CN towards \waql, emission from CN ($N=1-0, 2-1$) has been detected before with the IRAM-30m telescope \citep{bachiller1997_cn}, but at lower signal-to-noise ratios than attained in the APEX data. Using Kelvin-to-Jansky conversion factors of 38\,Jy/K for SEPIA/B5 \citep[][and references therein]{debeck2018_rdor} and 7.2\,Jy/K for the IRAM observations, in combination with a main-beam efficiency $\eta_{\rm{MB}}$ of 0.40 for the IRAM observations \citep{bachiller1997_cn}, we find that the peak fluxes for the strongest $N=2-1$ features in Fig.~\ref{fig:CN} reach $\approx3.8-4.6$\,Jy, whereas those measured by \citet{bachiller1997_cn} reach values of only $\approx1.2-1.4$\,Jy. Without performing any further modelling, we hypothesise here that stellar variability might be responsible for this factor of $\approx3$ difference in intensity, based on the conclusion by \citet{bachiller1997_cn} that radiative pumping through optical and near-IR bands likely dominates the CN excitation (calibration and/or pointing errors in the IRAM data cannot be excluded). The radiative-transfer models presented by \citet{danilovich2014} for \waql show an increase of CN at the photodissociation radius of HCN, in agreement with CN being produced as a photodissociation product of HCN. However, those models are only constrained by the $N=1-0$ and $N=2-1$ transitions and do not necessarily trace regions closer to the star or any possible variability of the emission over time.

\subsubsection{New detections: SiC$_2$, SiN, C$_2$H, HC$_3$N \label{sect:newdetections}}
\paragraph{SiC$_2$}
We detect fourteen emission lines pertaining to SiC$_2$ (Fig.~\ref{fig:SiC2}). We measure similar intensities for, for example,  the $9_{4, 6}- 8_{ 4, 5}$	(213.2\,GHz) and $ 9_{4, 5}- 8_{ 4, 4}$ (213.3\,GHz)  lines and for the $11_{4,8}-10_{4,7}$ (261.2\,GHz) and $11_{4,7}-10_{4,6}$ (261.5\,GHz) lines. This is in line with the two transitions in each pair having almost identical intrinsic strengths and upper-level energies. The pairwise match in intensities confirms the identification of SiC$_2$ in our spectra.  

We construct a rotational temperature diagram for SiC$_2$ based on the measured emission of twelve out of fourteen detected transitions (Fig.~\ref{fig:sic2_rotdiagram}). We leave out  two transitions: $11_{2,10}-10_{2,9}$ at 254.981\,GHz because of the large uncertainty owing to very high rms noise in this spectral region and $12_{0,12}-11_{0,11}$ at 261.990\,GHz which is blended with a much stronger signal from C$_2$H. We assume equal integrated intensity for the fully blended emission of the $10_{6,5}-9_{6,4}$ and $10_{6,4}-9_{6,3}$ transitions as they have identical transition probabilities (Einstein-$A$ coefficients). Under the assumption of optically thin emission excited under LTE conditions, we find a rotational temperature $T_{\rm rot}=76\pm28$\,K. We find a source-averaged column density of $(2.1\pm0.9)\times10^{14}$\,cm$^{-2}$ and an abundance SiC$_2$/H$_2\approx5\times10^{-7}$  when assuming that the emission arises in a spherical shell located at $1-2$\arcsec\/ in radius, i.e. at a distance $\sim400-800$\,AU from the star. The assumption of a peak in the SiC$_2$ abundance at roughly $600$\,AU from the star is based on (1) the fact that one rotational temperature seems to represent all detected emission in our data reasonably well, (2) the location in the CSE where the derived rotational temperature is reached according to the temperature profile presented by \citet{danilovich2014} is roughly 600\,AU, and (3) the fact that SiC$_2$ is also observed in a shell around \irc \citep{velillaprieto2019}. The SiC$_2$ emission towards \irc, however, is located in a shell at $\sim14-16$\arcsec\/ from the central star, that is at roughly three times larger physical separation from the central star compared to what we assume for \waql.  Assuming the equivalent spatial distribution for SiC$_2$ in \waql's CSE, we obtain a slightly higher abundance SiC$_2$/H$_2\approx8\times10^{-7}$, which is still well within the uncertainties of our simple approach here\footnote{The uncertainties on the abundance estimate stem from multiple simplifying assumptions, including optically thin emission, LTE excitation, and the match between rotational temperature and excitation temperature. The uncertainties therefore amount to more than the numerical uncertainties quoted for $T_{\mathrm{rot}}$ and $N_{\mathrm{col}}$, which are based only on uncertainties introduced by the data quality.}. The difference in spatial occurrence of SiC$_2$ could be a consequence of the roughly three times higher mass-loss rate of \irc, and hence  the correspondingly higher densities in the CSE. Comparing our estimated abundance to the results presented by \citet[][see their Fig.~6]{massalkhi2018_sic2} for a sample of 25 C-rich envelopes, we find that the combination of the SiC$_2$ abundance and wind density $\dot{M}/v_{\infty}$ of \waql fits well within the distribution of the C-rich envelopes, with \waql's SiC$_2$ abundance being on the lower end of the presented sample. Accounting for the uncertainties inherent to our approach, we conclude that this result indicates a possible similarity in chemistry between the CSE of the S-type AGB star \waql and C-type AGB stars.

\begin{figure*}
\centering
\includegraphics[trim={0 0 0 8.5cm},clip,width=0.8\linewidth]{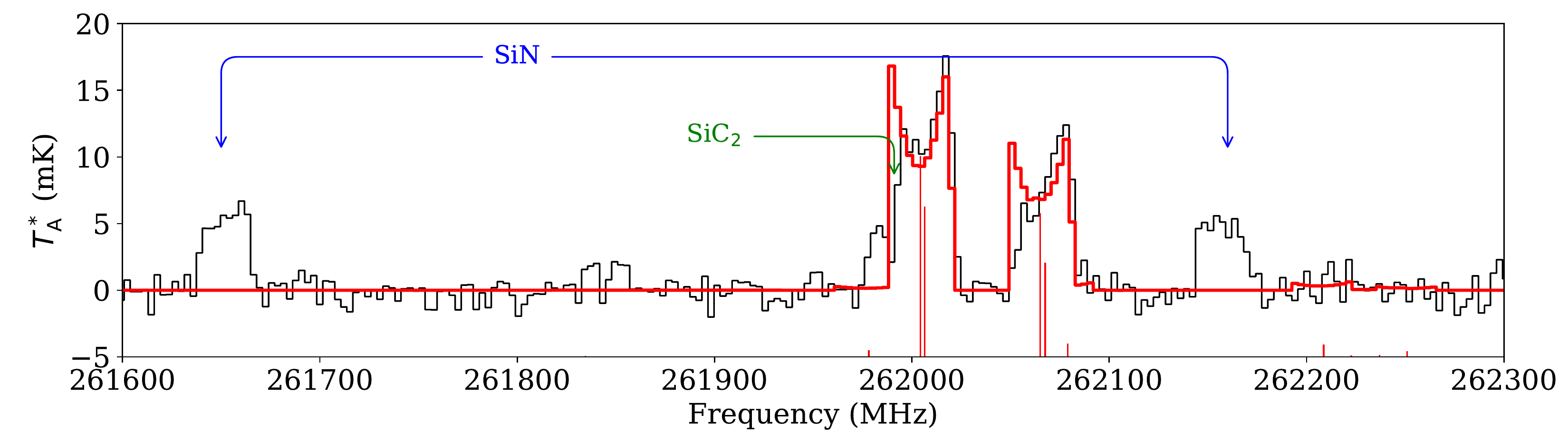} 
\caption{C$_2$H and SiN emission. High-signal-to-noise ratio detection of C$_2$H\,($3-2$) in the PI230 data, plotted at 3.0\,MHz ($\approx3.5$\,\kms) spectral resolution. Vertical red lines indicate the position and relative strength of the hyperfine structure components of the rotational transition of C$_2$H. In red, we overplot a synthetic spectrum of C$_2$H\,($3-2$) using  a "shell" profile as described in the GILDAS/CLASS manual, fitted to the data by eye, using the intrinsic component strengths. In blue we indicate the location of the strongest components of the SiN\,($N=6-5$) transition; see Fig.~\ref{fig:SiN} for all detected SiN emission. We mark the rest frequency of SiC$_2$\,($12_{0,12}-11_{0,11}$) in green. \label{fig:cch}}
\end{figure*}

\paragraph{SiN}
We detect emission from the strongest components of the SiN\,($N=5-4$) transition at 218.01\,GHz and 218.51\,GHz and of $N=6-5$ at 261.65\,GHz and 262.15\,GHz (Figs.~\ref{fig:cch} and \ref{fig:SiN}; these components have $\Delta F=\Delta J$). Assuming similar line strengths, the non-detection of the $N=4-3$ lines at 174.36\,GHz and 174.86\,GHz is in line with the much higher rms-noise in that region. The hyperfine structure of the SiN transition is not resolved and has a very limited contribution of only about 1\,MHz to the broadening of the overall profile. 

We can directly compare our APEX observations of SiN\,($N=6-5$) towards \waql to the NRAO 12\,m telescope observations of SiN\,($N=6-5$) towards \irc \citep{turner1992}. Scaling the emission for mass-loss rate and distance \citep[\waql: \mdot$=3\times10^{-6}$\,\msunyr, $d=395$\,pc; \irc: \mdot$=2\times10^{-5}$\,\msunyr, $d=130$\,pc; ][]{brunner2018_waql_alma,agundez2012_innerlayers_irc10216,menschikov2001}  we find that the SiN emission in \waql is roughly five times as bright as the SiN emission in \irc (for $N=6-5$). Assuming optically thin emission and the same excitation conditions for SiN in both CSEs, this would imply (SiN/H$_2$)$_{\mathrm{\waql}} \approx 5\times$(SiN/H$_2$)$_{\mathrm{\irc}} = 4\times10^{-8}$ \citep{agundez2009_phd}. 
\\

SiN and SiC$_2$ have both been detected towards the CSEs of C-type stars, but we are not aware of any reported detections of either molecule towards S-type or M-type AGB stars. Based on the angular resolution of 24--29\arcsec\/ obtained in the APEX observations of SiC$_2$ and SiN we cannot determine whether these molecules are present in the extended CSE of \waql or only close to the star, but from the width of the line profiles we find that both the SiC$_2$ and SiN emitting gas reach the wind's terminal velocity  $v_{\infty}\approx16.5$\,\kms \citep{danilovich2014}, suggesting that both molecules are present beyond the wind-acceleration zone. Furthermore, in the case of SiC$_2$, we determine a single rotational temperature low enough to be indicative of the molecule's presence in the outer parts of the CSE \citep[][see their Figs. 3 and 4]{danilovich2014}.

\begin{figure*}
\centering
\includegraphics[width=0.85\linewidth]{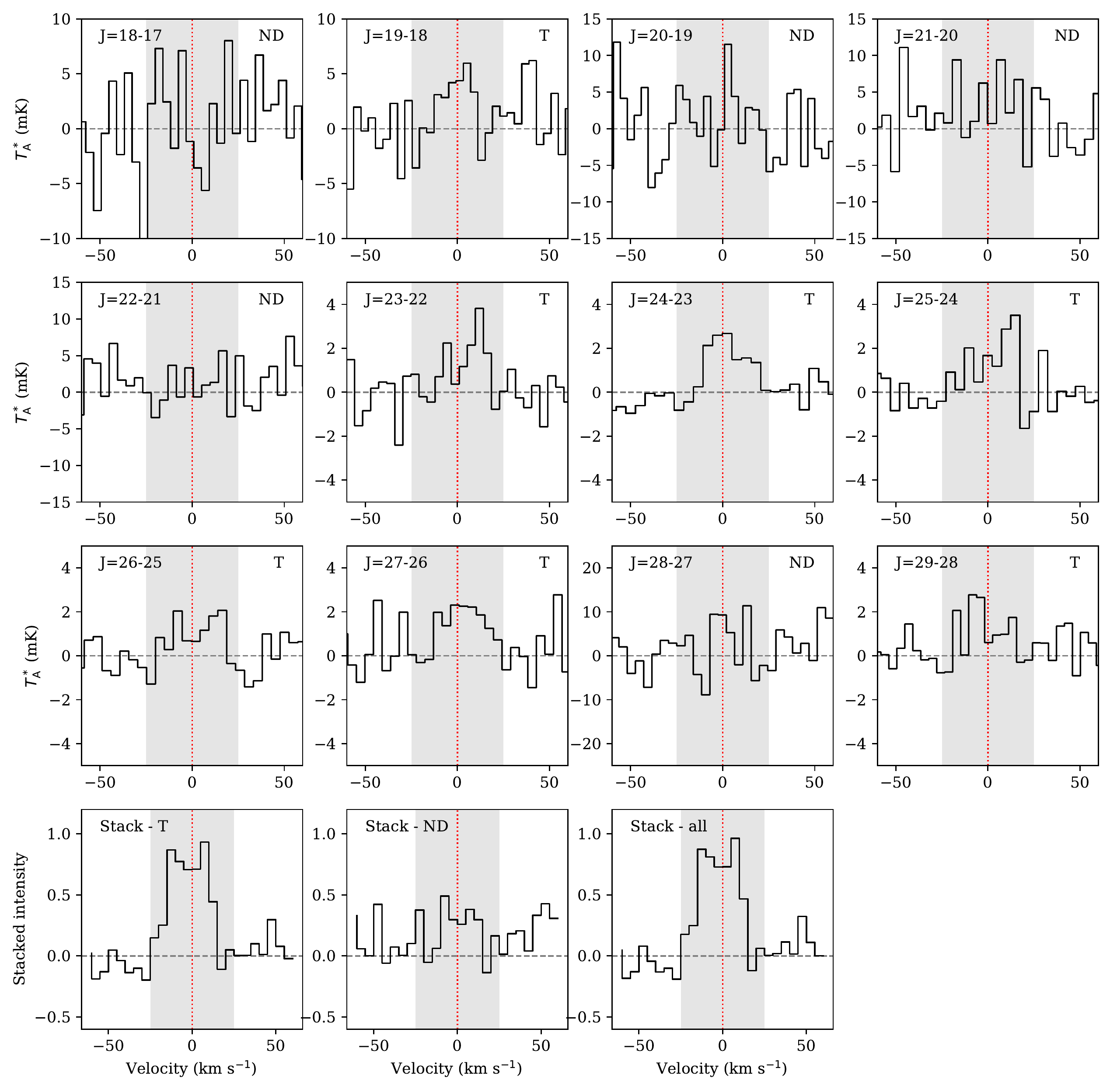} 
\caption{Detection of HC$_3$N emission.  \textit{Top three rows:} the spectra extracted around all frequencies of HC$_3$N rotational transitions in the vibrational ground state $v=0$ covered by our observations. Shaded areas indicate a velocity range $[-25,25]$\,\kms around the transition's rest frequency, vertical red lines correspond to the rest frequency centred to a velocity of 0\,\kms. All panels are labelled with the transition and whether the line is considered tentatively detected (T) or not detected (ND). \textit{Bottom row:} the result of stacking the emission for selected transitions, aligned in velocity space: \textit{(left) } tentatively identified transitions (marked T), \textit{(middle)} transitions not detected (marked ND), \textit{(right)} all twelve transitions shown in the top three rows. Note that the intensity scale is normalised to the peak and has no physical meaning as such. \label{fig:HC3N}}
\end{figure*}

\paragraph{C$_2$H}
We tentatively detect emission from C$_2$H in the  $N=2-1$ transition at $\sim$174.7\,GHz, and detect emission at high signal-to-noise ratio for $N=3-2$ at $\sim$262.0\,GHz (Fig.~\ref{fig:cch}). We produce a synthetic spectrum for the $N=3-2$ transition assuming for each hyperfine component a "shell" profile, as described in the GILDAS/CLASS manual, and using the tabulated intrinsic relative strengths and fit the sum of them by-eye to our data (see Fig.~\ref{fig:cch}). We find that the measured relative strengths of the hyperfine structure components agree very well with the intrinsic relative strengths tabulated in the CDMS \citep{muller2000_cch}. 

The line profiles of the doublet components of C$_2$H\,($N=3-2$) are double-peaked, but we note that the peaks corresponding to the extreme velocities of the doublet components are not equal in strength, the blue-shifted peaks being much stronger than the red-shifted peaks in both components.  The strong difference between the blue and red sides of the profiles hints at an asymmetry in the CSE with more C$_2$H gas excited in the part of the CSE moving towards the observer. We cannot assess whether this would be caused by differences in excitation or in abundance distribution. Both of these could be a consequence of, for example, clumpiness in the CSE. 

We also note that roughly the opposite is true in the C$_2$H spectra towards \irc, where the red-shifted peaks are stronger than the blue-shifted peaks in the transitions $1-0,\dots,7-6$ presented by  \citet{debeck2012_cch}. In the case of \irc, the difference is likely a consequence of optical-depth effects, where only the outermost, and hence colder, C$_2$H gas is visible in the front of the CSE, whereas we only see the emission from the warmer, inner C$_2$H gas from the part of the CSE that moves away from the observer. In the case of \irc, we know that the C$_2$H emission is spread over a thin, circular shell around the star. In the case of \waql we lack direct information on the spatial distribution of the excited C$_2$H gas, but the double-peaked spectrum suggests emission from a shell.

Scaling the measured C$_2$H line emission for \irc to what one could expect at the distance of \waql and assuming for simplicity that the emission is excited in a similar shell-like region around the star and that the line intensity scales linearly with \mdot, we expect the strongest components to peak at $\sim10-15$\,mK for $N=2-1$ and $N=3-2$ in our APEX data. Since the observed lines peak at $\sim 5-15$\,mK, we suggest that the abundance of C$_2$H in the CSE of \waql is roughly a factor 2 lower than in the CSE of \irc, i.e. (C$_2$H/H$_2$)$_{\mathrm{\waql}} \approx 0.5\times$(C$_2$H/H$_2$)$_{\mathrm{\irc}}=1\times10^{-5}$ \citep{debeck2012_cch}. \citet{cernicharo2014_timevariability} note that the strength of the C$_2$H emission in \irc is highly variable with time, as a consequence of the strong effect of radiative pumping in the molecular excitation and the changes in the stellar radiation field owing to pulsations. However, since we consider here a low-$N$ rotational transition this variability is likely limited to within the observational uncertainties (see their Fig.~2).

We do not detect any emission from the longer-chain radicals C$_4$H or C$_6$H at the current sensitivity.

\paragraph{HC$_3$N}
We tentatively detect emission in multiple rotational transitions from HC$_3$N. Figure~\ref{fig:HC3N} shows an overview of the spectral ranges where rotational transitions of HC$_3$N are covered by our data. Some transitions are not detected (marked ND) and some are tentatively detected (marked T). We also show the result of stacking the data, aligned in velocity space, weighted with the ratio of the statistical weights $2J+1$ of the upper levels of the respective transitions $J\rightarrow J-1$ and the rms noise in the spectrum close to the transition's rest frequency. We performed this stacking procedure for three sets of lines; we obtain $S/N>5$ when stacking the tentative detections, do not find significant structure when stacking the non-detections, and find that the tentative detections weigh strong enough to give rise to line-like structure also in the total stacked profile. This is a direct result of the higher rms noise in the parts of the spectrum with the non-detections, making these less significant.

Based on the multiple tentative detections and the results from the stacking procedure we are confident that we detect HC$_3$N in the CSE of \waql. As far as we know, it is the first time such a carbon-chain molecule is seen in the CSE of an S-type star and it has not yet been detected in the CSE of an M-type AGB star.  Comparing the line peak intensities to what \citet{tenenbaum2010} report for \irc, we find that the emission is weaker in \waql by a factor $10-15$, accounting for the difference in distance and mass-loss rate. This implies that HC$_3$N is roughly one to two orders of magnitude less abundant in the CSE of \waql than around \irc and leads to a first estimate of a fractional abundance on the order $10^{-7}-10^{-8}$ \citep{agundez2009_phd}.

\begin{figure}[htb!]
\centering
\subfigure[HN$^{13}$C \label{fig:HN13C}]{\includegraphics[width=.96\linewidth]{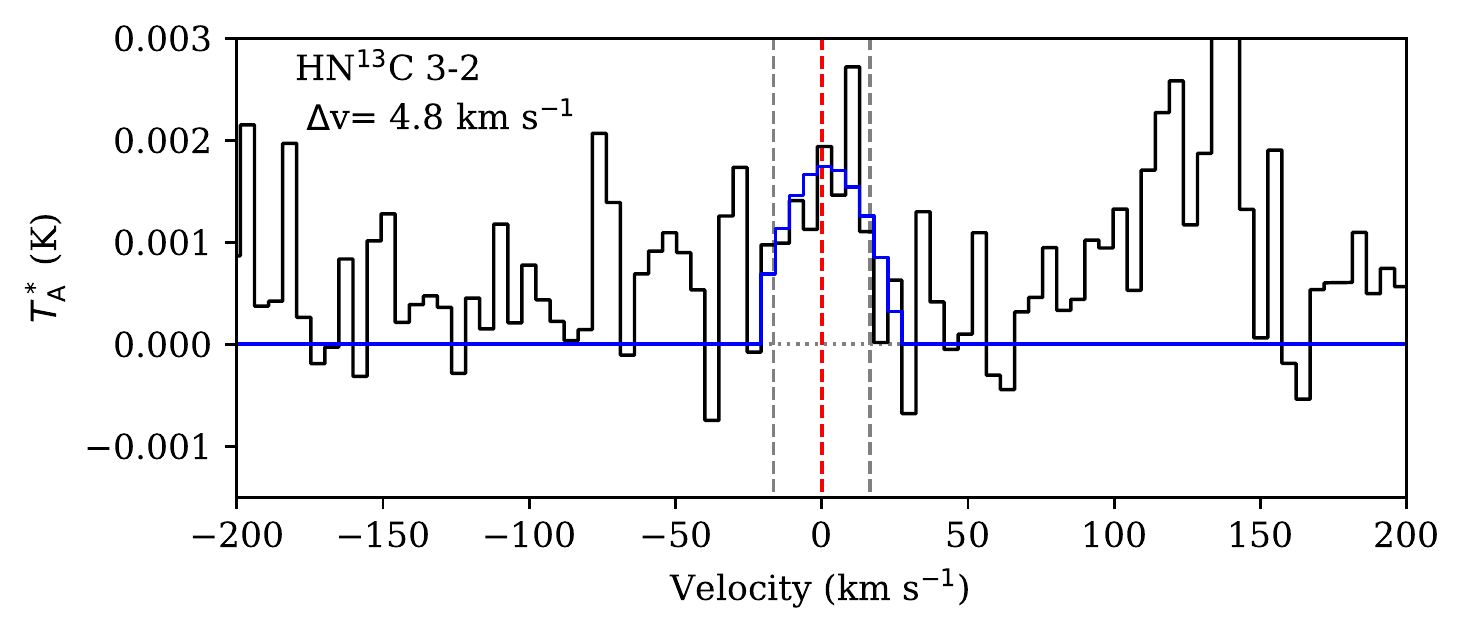}} 
\subfigure[NaCl  \label{fig:NaCl}]{\includegraphics[width=.96\linewidth]{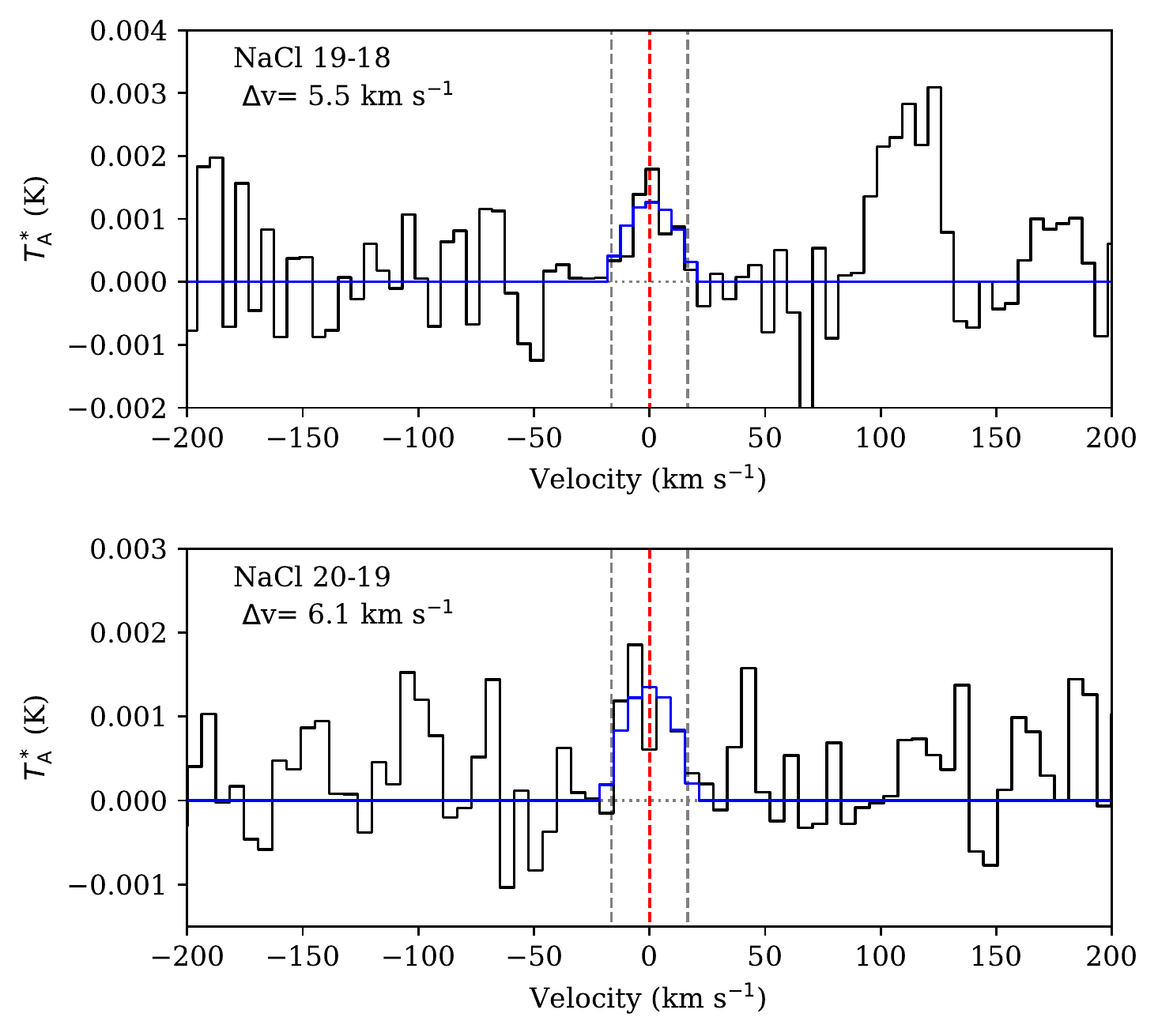}} 
\subfigure[HCO$^+$ \label{fig:hco+}]{\includegraphics[width=.96\linewidth]{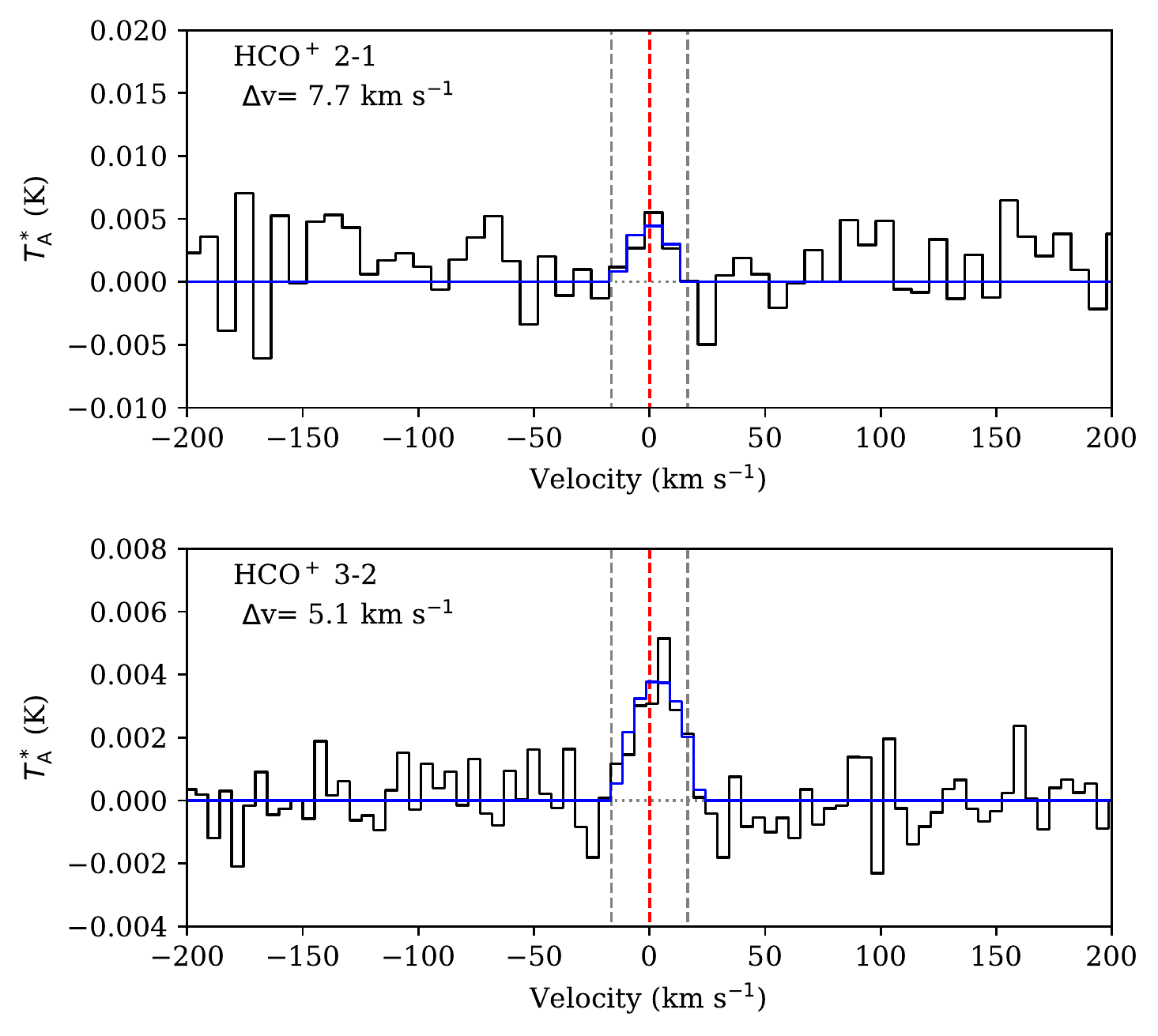}} 
\caption{Tentative detections. Red vertical lines correspond to the transitions' rest frequencies; grey vertical lines indicate the 16.5\,\kms expansion velocity of \waql's CSE; blue fits were produced using the "shell" profile function as defined in the GILDAS/CLASS package. The velocity resolution at which the spectra are presented is listed in each panel.}
\end{figure}

\subsubsection{Tentative detections}
\paragraph{HNC}
We tentatively detect emission from HN$^{13}$C\,($J=3-2$) at 261.3\,GHz at a peak-S/N of $\approx$3, see Fig.~\ref{fig:HN13C}. For this tentative detection, and those of NaCl and HCO$^{+}$ discussed below, we quote S/N values of line profile fits based on the "shell" profile as defined in the GILDAS/CLASS manual, using the rms noise in the line-free part of the spectra at $[-200;-50]$\,\kms. 

We do not detect emission from HN$^{13}$C\,($J=2-1$) at 174.2\,GHz or HNC\,($J=2-1$) at 181.3\,GHz, in line with the sensitivity reached in our data set at the relevant frequencies.  Assuming optically thin emission from both H$^{13}$CN and HN$^{13}$C we get an abundance ratio HN$^{13}$C/H$^{13}$CN of the order $0.01-0.1$. Values reported in the literature for the abundance ratio HNC/HCN are anywhere in the range 0.001 and 0.1 for  the CSEs of the C-type AGB star \irc and the supergiant \vycma \citep{johansson1984,daniel2012_hnc,ziurys2009_vycma_Cchem}. For the remainder of the paper, we assume that $\mbox{HNC/H}_2 = 0.01\times\mbox{HCN/H}_2 = 3.1\times10^{-8}$ \citep{danilovich2014} with one order of magnitude uncertainty in both directions, in agreement with our HN$^{13}$C measurements and the above literature values.

\paragraph{NaCl}
We tentatively detect NaCl\,($J=19-18,20-19$) at 247.2\,GHz and  260.2\,GHz, see Fig.~\ref{fig:NaCl}. We obtain peak-signal-to-noise ratios of $\approx2$ and integrated-signal-to-noise ratios of $\approx4$ for both transitions. The non-detections of NaCl emission in the SEPIA-B5 data are consistent with the higher rms noise in those observations.  Based on a comparison to the \iktau spectrum presented by \citet{velillaprieto2017_iktau_iram} we estimate that $(\mbox{NaCl/H}_2)_{\mathrm{\waql}}\approx 0.1 (\mbox{NaCl/H}_2)_{\mathrm{\iktau}}$; based on the NaCl detections towards \irc \citep{agundez2012_innerlayers_irc10216} we estimate that $(\mbox{NaCl/H}_2)_{\mathrm{\waql}}\approx 4 (\mbox{NaCl/H}_2)_{\mathrm{\irc}}$. This leads to an estimate for $(\mbox{NaCl/H}_2)_{\mathrm{\waql}}$ on the order of a few $10^{-8}$.The current data do not provide reliable information on which part of the CSE this emission originates from and follow-up observations at higher sensitivity are needed to draw any firm conclusions on the presence, abundance, and/or location of NaCl in the CSE. NaCl emission has been detected around both M-type and C-type AGB stars \citep[e.g.,][]{agundez2012_innerlayers_irc10216,debeck2015_AGBconfproceedings,debeck2018_rdor}, but we are not aware of any reported detection towards S-type stars.

\paragraph{HCO$^+$}
We tentatively detect emission from HCO$^+$\,($J=3-2$) at 267.6\,GHz at a peak-S/N of $\approx$4, see Fig.~\ref{fig:hco+}. There clearly is emission at this position in the spectrum, but we cannot firmly claim that HCO$^+$ is the carrier since we have no other detected transition towards \waql to confirm this. However, based on the tentative identification, we also present the spectrum at 178.4\,GHz, where the HCO$^{+}$\,($J=2-1$) transition lies. We cannot claim a detection, but do observe structure in the spectrum suggestive of a low-intensity emission line at the correct position. A confirmation will require sensitive follow-up observations of multiple transitions.

HCO$^+$ has been detected in the CSEs of the M-type AGB stars \iktau, \txcam, and \whya \citep{pulliam2011_hcoplus}, in the bipolar lobes around the peculiar OH/IR star \object{OH~231.8+4.2} \citep{sanchezcontreras2015_oh231}, and around the supergiants \vycma and \nmlcyg \citep{ziurys2007,ziurys2009_vycma_Cchem,pulliam2011_hcoplus}. All of these CSEs showcase an oxygen-rich chemistry.   Typical abundances HCO$^{+}$/H$_2$ range around $10^{-8}-10^{-7}$. HCO$^+$ has also been detected in the CSE of \irc  \citep[][and references therein]{agundez2006_Ochemistry_irc10216,pulliam2011_hcoplus}, but its abundance in that case is as low as a few $10^{-9}$. \irc is currently the only C-type AGB star for which such a detection has been reported. The emission (suspected) from HCO$^+$\,($3-2$) towards \waql is a factor of roughly 3, 7, and 40 stronger than those reported by \citet{pulliam2011_hcoplus} towards \txcam, \iktau, and \irc, respectively, accounting for the mass-loss rates, distances, and difference in beam size between APEX and the SMT.  We assume for the remainder of the paper that $(\mbox{HCO}^+\mbox{/H}_2)_{\mathrm{\waql}}\lesssim  7\times (\mbox{HCO}^+\mbox{/H}_2)_{\mathrm{\iktau}}=7.0\times10^{-8}$ \citep{velillaprieto2017_iktau_iram}.

\subsubsection{Upper limits}
\paragraph{SO, SO$_2$}
We do not detect emission from SO or SO$_2$ in our spectra of \waql, whereas these are dominant in the spectra of the M-type AGB stars \rdor and \iktau \citep{debeck2018_rdor,velillaprieto2017_iktau_iram}. From a comparison with scaled spectra of \rdor  and \iktau in the SEPIA/B5 range \citep[][and De Beck et al., \emph{in prep.}]{debeck2018_rdor,velillaprieto2017_iktau_iram}, accounting for the differences in distance and mass-loss rate \citep[$d_{\mathrm{\rdor}}=59$\,pc, $\dot{M}_{\mathrm{\rdor}}=2\times10^{-7}$\,\msunyr, $d_{\mathrm{\iktau}}=250$\,pc, $\dot{M}_{\mathrm{\iktau}}=8\times10^{-6}$\,\msunyr;][]{maercker2016_water,decin2010_nlte}, we set \rdor's SO$_2$ abundance ($5\times10^{-6}$) and \iktau's SO abundance ($1\times10^{-6}$) as upper limits for the respective abundances in \waql's CSE. We discuss these differences and possible implications for the sulphur chemistry in Sect.~\ref{sect:discussion}.

\paragraph{H$_2$S}
We do not detect emission from H$_2$S in our spectra, although we obtained data at slightly higher sensitivity than \citet{danilovich2017_h2s}. From a direct comparison to the line emission from \iktau reported by \citet{danilovich2017_h2s}, we set an upper limit $(\mbox{H$_2$S/H}_2)_{\mathrm{\waql}} < \frac{1}{3}\times (\mbox{H$_2$S/H}_2)_{\mathrm{\iktau}}=5\times10^{-7}$ \citep{danilovich2017_h2s}.

\paragraph{PN, PO}
Similarly, we derive an upper limit $(\mbox{PN/H}_2)_{\mathrm{\waql}} <  (\mbox{PN/H}_2)_{\mathrm{\iktau}} = 3\times10^{-7}$ based on the comparison of our observations to the emission spectrum of PN\,($J=5-4$) presented by \citet{velillaprieto2017_iktau_iram}. A comparison to the spectra presented by \citet{milam2008_phosphorus} for \irc does not set additional constraints as the scaled intensity lies far below the sensitivity reached in our observations. Unfortunately, we cannot derive an upper limit for the PO abundance towards \waql, as the sensitivity of the  \iktau spectrum obtained with SEPIA/B5 (scaled for $d$ and \mdot) is superior by a factor $\sim$8 to that of the \waql spectrum and the PO transition at 196\,GHz remains undetected in both spectra.

\subsubsection{Unidentified lines}
We find only two unidentified features in this rather broad spectral coverage. Figure~\ref{fig:ulines} shows emission features in our spectra that we cannot identify with any known carriers. We can rule out contamination of the spectrum through the image sideband for all of these.  Two of the three features lie immediately adjacent to each other, are similar in intensity, and are located close to the $^{29}$SiS\,($v=3,J=14-13$) line. With the current sensitivity of the data we cannot rule out that these two features (at $249393\pm4$\,MHz and $\approx249410$\,MHz) are actually part of the same emission line, centred at $249402\pm4$\,MHz.

Another unidentified feature is centred at $267073\pm4$\,MHz and seems double-peaked. We cannot identify any possible carrier within 15\,MHz of the presumed central frequency.

\begin{table}
\caption{Isotopologue ratios retrieved from frequency corrected line-intensity ratios $R_{\mathrm{c}}$, as described in the text. Note that the reported errors are formal errors, derived from the errors on the respective integrated line strengths. \label{tbl:isotopes}}
\small
\begin{tabular}{ccr|cc}
\hline\hline\\[-2ex]													
Isotopologue	&	Transition	&		$R_\mathrm{c}$		&	\multicolumn{2}{c}{Solar} 	\\
Ratio				&					&									&	\multicolumn{2}{c}{isotope ratio} 	\\

\hline\\[-2ex]													
$^{12}$CO/$^{13}$CO	&	$2-1$	&	$	12.3	\pm	0.5	$	&$^{12}$C/$^{13}$C	&	89\\
H$^{12}$CN/H$^{13}$CN	&	$2-1$	&	$	5.1	\pm	0.9	$	&		\\
$^{12}$CS/$^{13}$CS		&	$5-4$	&	$	21.7	\pm	3.4	$	&		\\
$^{12}$CN/$^{13}$CN 	&$2-1$	& $16.7 \pm 4.2$ & \\
\hline\\[-2ex]

$^{28}$SiO/$^{29}$SiO	&	$4-3$	&	$	7.5	\pm	1.3	$	&	$^{28}$Si/$^{29}$Si	&	20	\\
									&	$5-4$	&	$	7.4	\pm	0.6	$	&		\\
$^{28}$SiS/$^{29}$SiS	&	$12-11$	&	$	20.4	\pm	6.8	$	&		\\
									&	$14-13$	&	$	12.8	\pm	3.1	$	&		\\

$^{28}$SiO/$^{30}$SiO	&	$4-3$	&	$	8.4	\pm	1.1	$	&	$^{28}$Si/$^{30}$Si		&30	\\
									&	$6-5$	&	$	9.3	\pm	1.3	$	&		\\
$^{28}$SiS/$^{30}$SiS	&	$12-11$	&	$	14.5	\pm	1.9	$	&		\\
									&	$13-12$	&	$	12.3	\pm	3.1	$	&		\\
									&	$14-13$	&	$	28.2	\pm	8.0	$	&		\\
$^{29}$SiO/$^{30}$SiO	&	$4-3$	&	$	1.1	\pm	0.2	$	&	$^{29}$Si/$^{30}$Si		&1.5	\\
$^{29}$SiS/$^{30}$SiS	&	$12-11$	&$	0.7	\pm	0.2	$	&		\\
									&	$14-13$	&$	2.2	\pm	0.5	$	&		\\
									&	$15-14$	&	$	0.3	\pm	0.1	$	&		\\
\hline\\[-2ex]

Si$^{32}$S/Si$^{34}$S	&	$14-13$	&	$	10.6	\pm	2.6	$	&	$^{32}$S/$^{34}$S	&22	\\
\hline\\[-2ex]
C$^{17}$O/C$^{18}$O	&	$2-1$	&	$	1.0	\pm	0.4	$	&	$^{17}$O/$^{18}$O	&0.19	\\
C$^{16}$O/C$^{18}$O	&	$2-1$	&	$	355.7	\pm	112.0	$	&	$^{16}$O/$^{18}$O	&	499	\\
Si$^{16}$O/Si$^{17}$O	&	$6-5$	&	$	74	.3   \pm	5.4	$	&	$^{16}$O/$^{17}$O	&	2632	\\
\hline													
\hline					
\end{tabular}\normalsize										
\tablefoot{We disregard $^{30}$SiO($J=5-4$) at 211.853\,GHz and Si$^{34}$S\,($J=12-11$)  at 211.854\,GHz in this table as they are fully blended. Solar values are taken from \citet{asplund2009_solarabundances}.}
\end{table}

\subsection{Isotopes}\label{sect:isotopes}
Isotopic ratios can ideally be used to help constrain the initial mass of the star and its current evolutionary status. We present estimates of isotopic abundance ratios in Table~\ref{tbl:isotopes}. These are assumed to be represented by the abundance ratios $R_{\mathrm{a}/\mathrm{b}}$ of isotopologues a and b, estimated according to 
\begin{equation}
R_{\mathrm{a}/\mathrm{b}} = \frac{I_{\mathrm{a}(J-J^{\prime})}}{I_{\mathrm{b}(J-J^{\prime})}} \times \left(  \frac{\nu_{\mathrm{a}(J-J^{\prime})}}{ \nu_{\mathrm{b}(J-J^{\prime})}} \right)^{-3},
\end{equation}
using the integrated line strengths $I$ of transitions $J-J^{\prime}$ for isotopologues a and b and taking into account the difference in intrinsic line strengths between the isotopologue transitions and telescope beamwidths at the rest frequencies. This simple method assumes that the same transition $J-J^{\prime}$ for two isotopologues is excited under similar conditions.

\paragraph{Carbon}
From radiative transfer modelling of multiple emission lines from both \twco and \thco \citet{ramstedt2014_12co13co} and  \citet{danilovich2014} derived \twth of 26 and 29, respectively. We do not attempt such modelling here, but it is clear that the CO, HCN, CS, and CN emission are affected by optical-depth effects since the relevant line intensity ratios are much lower than this (see Table~\ref{tbl:isotopes}). 

\paragraph{Oxygen}
The detections of C$^{17}$O\,($J=2-1$) and C$^{18}$O\,($J=2-1$) lead to an estimate of $\mathrm{C}^{17}\mathrm{O}/\mathrm{C}^{18}\mathrm{O}=1.0 \pm 0.4$. We note that the C$^{18}$O emission in our data appears blended with an unidentified source of emission, increasing the uncertainty on this estimate. This component is not visible in the spectra presented by \citet{denutte2016_17o18o} owing to their higher noise. We detect emission from Si$^{17}$O\,($J=6-5$), but not from Si$^{18}$O\,($J=6-5$). This non-detection is in line with the achieved sensitivity, when considering the above results.

\paragraph{Silicon}
We detect emission from multiple transitions of $^{28}$SiO,  $^{29}$SiO, $^{30}$SiO and  $^{28}$SiS,  $^{29}$SiS, $^{30}$SiS; see Figs.~\ref{fig:SiO} and \ref{fig:SiS}. However, based on the ratios sampled by  the different transitions, it is clear that optical-depth effects need to be accounted for in order to derive reliable isotopic ratios. For example, we find $^{28}\mathrm{Si}/^{29}\mathrm{Si}$ isotopologue line intensity ratios spanning the range $7-20$ and $^{28}\mathrm{Si}/^{30}\mathrm{Si}$ spanning the range $8-28$. Detailed abundance modelling using non-LTE radiative transfer models is, however, beyond the scope of this paper. 

\paragraph{Sulphur}
We detect  Si$^{32}$S and Si$^{34}$S, and  C$^{32}$S and C$^{34}$S. We do not detect any carriers of $^{33}$S, the least abundant sulphur isotope. The uneven sampling of the spectrum of the different isotopologues prevents estimates of the isotopic ratios without detailed modelling.

\paragraph{Initial stellar mass}
\citet{danilovich2015_waql_binary} constrained the current mass of \waql to be in the range $1.04-3$\,\msun. \citet{denutte2016_17o18o} derived an initial mass $M_{\mathrm{i}} = 1.59\pm0.08$\,\msun based on their analysis of C$^{17}$O and C$^{18}$O emission. Using our observations we derive an initial-mass estimate of $M_{\mathrm{i}}=1.6\pm0.2$\,\msun. Both our estimate and the one of \citeauthor{denutte2016_17o18o} are based only on the relative emission strengths of the $J=2-1$ transitions of C$^{17}$O and C$^{18}$O. The derived initial mass appears consistent with the fact that evolutionary models predict only stars with initial mass above $\sim$1.5\,\msun  \citep[e.g. ][]{hinkle2016_isotopicratios} to undergo third dredge-up events, a necessary process to bring \textit{s}-process elements to the surface and move the star from M-type to S-type.  \citet{hinkle2016_isotopicratios} report that their model for a star with $M_{\mathrm{i}}=1.6$\,\msun reaches a $\mathrm{C/O}=1$ when $^{12}$C/$^{13}$C reaches 28. The reported C/O and $^{12}$C/$^{13}$C and our derived initial mass hence all fit into one consistent picture of \waql as a star close to its transition from being oxygen-rich to becoming carbon-rich.

\section{Discussion}\label{sect:discussion}
Based on the classification scheme of \citet{keenan1980_Stype},  \citet{danilovich2015_waql_binary} reported a spectral type S6/6e and a $\mbox{C/O}=0.98$, making \waql an S-type star that is close to having a carbon-rich atmosphere. On the other hand, based on the presence of emission from SiN, SiC$_2$, C$_2$H, and HC$_3$N in the CSE, one may classify the content of \waql's CSE as (rather) carbon-rich. This apparent contrast requires further discussion. In this section we  compare the appearance of \waql's atmosphere and circumstellar gas and dust to what we know about other AGB stars.

\subsection{Atmosphere}
\citet{hony2009_S_ISO} reported the presence of HCN, C$_2$H$_2$, and C$_3$ absorption bands towards \waql in the mid-infrared ISO data, but did not discuss this. These molecules are typical for C-type atmospheres, but their formation is maybe possible also in atmospheres of stars with C/O\,$\lesssim1$ as a consequence of the non-equilibrium chemistry induced by shocks that travel through the atmosphere and can free atomic carbon due to the destruction of, mainly, CO. Although the models presented by \citet{cherchneff2006} indeed show an increase in C$_2$H$_2$ abundance in the presence of shocks compared to thermal equilibrium, they also show an approximately three orders of magnitude drop in the C$_2$H$_2$ abundance at 5\,\rstar when varying C/O from 1 down to 0.98 (see their Fig.~7). The predicted fractional abundance in the latter case, $\approx10^{-9}$, is still more than 5 orders of magnitude higher than for stars with $\mbox{C/O}<0.9$. For reference, \irc has a derived C$_2$H$_2$ abundance of $8\times10^{-5}$ in the inner CSE \citep[within 40 stellar radii;][and references therein]{fonfria2008}. The models presented by \citet{agundez2020_chemequilibrium} indicate that, even under the assumption of chemical equilibrium, the molecular gas around S-type stars with $\mbox{C/O}\approx1$ could resemble that around C-type stars quite closely, with similar abundances of, for example, C$_2$H$_2$ and HCN in the atmosphere for $\mbox{C/O}$ in the range $0.98-1.02$. The presence of C$_2$H$_2$ in the stellar atmosphere of \waql explains the presence of C$_2$H in its circumstellar environment, as the radical is the direct photodissociation product. The presence of both C$_2$H$_2$ and HCN  in the atmosphere can explain the presence of HC$_3$N in the CSE, according to the chemical models of  \citet{cordiner2009}.

\waql is a Mira star with a variability in the visual $\Delta V\approx 7\mbox{ mag}$. Such stars populate region C in the ($K-[12],[25]-[60]$) colour-colour diagrams presented by \citet{jorissen1998_sstars}. \citet{vaneck2017} point out that strong variability can significantly affect the estimated atmospheric parameters such as the effective temperature, surface gravity, C/O, and s-process enrichment [s/Fe]. \citet{keenan1980_Stype} report observations of \waql at only three different phases, making it difficult to assess this potential variability.  It is beyond the scope of this paper to quantify the possible effect of the photometric variability on the estimated C/O. However, we repeat that the absence of TiO lines and the presence of strong Na D lines put \waql close to the C-type classification, whereas the presence of ZrO bands points at \waql not being sufficiently enriched yet to be C-type.

\begin{figure*}
\includegraphics[width=\linewidth]{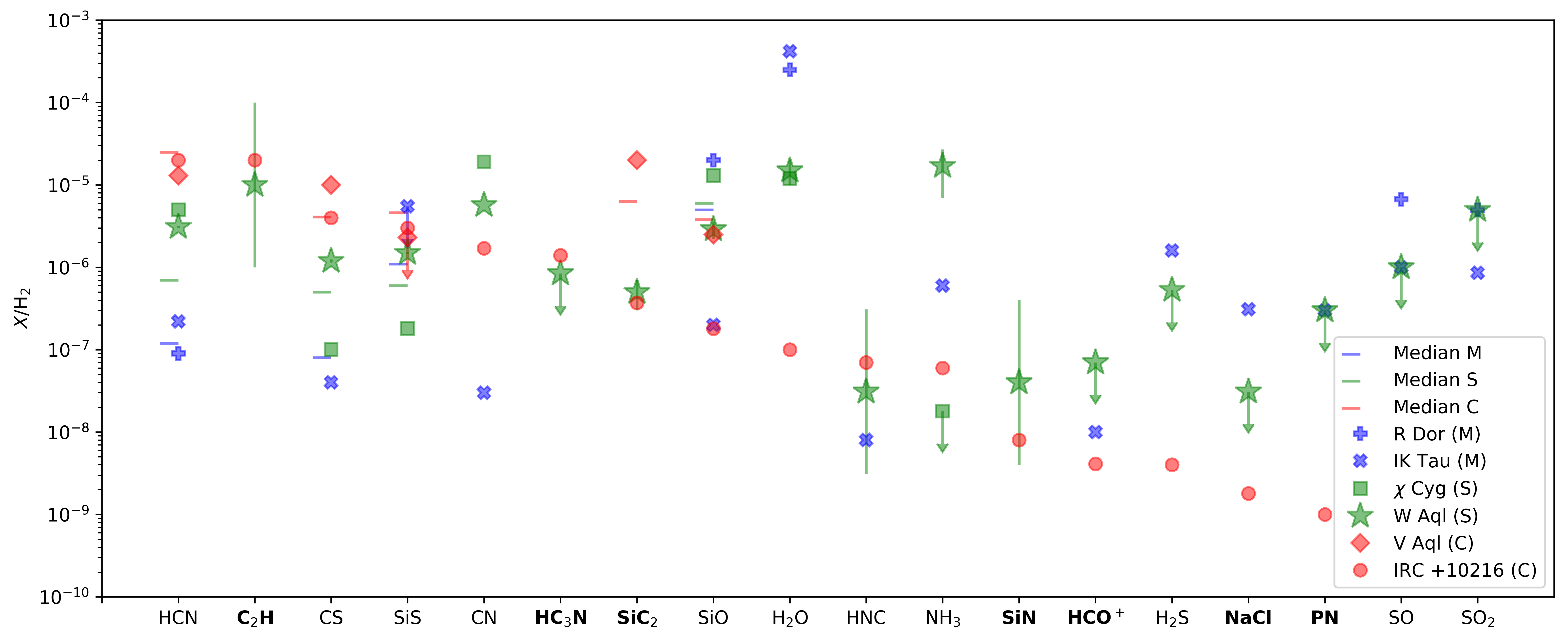} 
\caption{Peak molecular abundances in the CSEs of AGB stars of different chemical types: M-type stars \rdor and \iktau, S-type stars \chicyg and \waql, and C-type stars \vaql and \irc. We note that the high SiS abundance for \iktau is only representative of a geometrically small component centred on the star, after which depletion is very efficient and the abundance drops roughly 4 orders of magnitude. See text for further discussion and references.  We have included error bars only for \waql for the sake of visibility. The error bars correspond to the values specified in Table~\ref{tbl:compabundance} and we note that they are often smaller than the plotting symbol. Upper limits are indicated with downwards arrows. Molecules that were not  previously identified towards \waql are highlighted in boldface font on the horizontal axis. \label{fig:compabundance}}
\end{figure*}

\subsection{Circumstellar gas} 
Central to the discussion on how \waql's circumstellar chemistry compares to that of other AGB CSEs is Fig.~\ref{fig:compabundance}. We show estimated peak fractional abundances for CSEs around a small collection of AGB stars: the two M-type stars \rdor and \iktau, the two S-type stars \chicyg and \waql, and the two C-type stars \vaql and \irc. For all three chemical types the former star has a mass-loss rate in the range $0.1-0.7\times10^{-6}$\,\msunyr and the latter in the range $3.0-20\times10^{-6}$\,\msunyr. For a few molecules, we also show median values for different chemical types based on sample studies\footnote{We wish to remark, however, that some sample sizes are limited and reported values are in those cases likely biased towards high abundances. We refer to the papers in question for details on samples and biases.} and list the used peak fractional abundances and the references to the relevant studies in Table~\ref{tbl:compabundance}. 

\citet{danilovich2014} found that their derived circumstellar abundances of HCN, SiO, and H$_2$O for \waql are between those typical for M- and C-type AGB stars, from which they conclude that the abundances are consistent with an S-type classification. \citet{schoeier2013_hcn} reported that the HCN abundance in S-type CSEs indeed is intermediate between those in CSEs of M-type and C-type AGB stars. 

Contrary to the fact that \waql's CSE seems to have chemical properties intermediate between the M-type and C-type classes when looking at the \textit{usual suspects} such as CS, SiS, HCN, and SiO, we estimate high abundances of  C$_2$H, CN, SiC$_2$, and SiN, and likely also of HC$_3$N, and HNC. All of these molecules  are expected around C-type stars and are not seen around M-type stars. Given that ${\rm C/O}<1$ for S-type stars like \waql, one does not, at first sight, expect such a carbon-rich chemistry as we report in Sect.~\ref{sect:results} and show in Fig.~\ref{fig:compabundance}. For molecules where CSEs from M-type stars exhibit markedly higher abundances than those from C-type stars, for example H$_2$S, PN, PO, SO, and SO$_2$, we have no estimates or only upper limits for the case of \waql. From our results and the abundances reported in the literature and compiled in Fig.~\ref{fig:compabundance} it seems that \waql's CSE is closer to that of \irc in its gas-phase chemistry than to that of \iktau, whereas all three stars have roughly similar wind properties (expansion velocity and mass-loss rate). Only in the cases of H$_2$O and NH$_3$ are the abundances for \waql closer to those for \iktau than to those for \irc; for SiO, the reported abundance is exactly intermediate\footnote{We cannot make such a claim with certainty in the cases of HCO$^+$ and NaCl, since our detections are tentative and our abundance estimates very rough. Differences within an order of magnitude are hence not significant at this point.}. In the case of NH$_3$, \waql's abundance exceeds that of \iktau, removing it entirely from an intermediate status. We note, though, that the excitation of all three molecules (H$_2$O, NH$_3$, and SiO) is predominantly through the strongly variable radiation field, possibly introducing considerable uncertainties on the derived abundances (for all objects).

Currently, it is difficult to assess the differences or similarities between the CSE of \waql ($\mbox{C/O}=0.98$), on the one hand, and that of \chicyg \citep[$\mbox{C/O}\leq0.95$; based on the measurements and classification scheme presented by][]{keenan1980_Stype}, on the other hand, as broad, sensitive observations of \chicyg are currently missing. The chemical equilibrium models of \citet{agundez2020_chemequilibrium} provide support for an inner-wind chemistry around \waql closely resembling that of a C-type AGB star. However, the molecules we detect that are ``typical for C-rich CSEs'' are found in the outer CSE and there are currently no chemical models available which describe the chemical content of the CSEs of S-type AGB stars on that physical scale. Because of this and the lack of observational characterisation of CSEs of S-type stars with different $\mbox{C/O}$, we cannot yet assess in a straightforward way how (un)typical the CSE of \waql is. Chemical models must explore the impact of the atmospheric C/O on the circumstellar abundances.

\subsection{Dust}
\citet{hony2009_S_ISO} proposed the presence of non-stoichiometric silicates around S-type stars to explain the substantial spectral differences in the infrared between CSEs of S-type and M-type AGB stars, which is plausible since stars with $\mathrm{C/O}\sim1$ exhibit an increased Mg/SiO$_4$ abundance ratio, with less free oxygen available when dust forms. Additionally, they reported the presence of the 30\,$\mu$m feature, attributed to MgS, in the ISO spectrum of \waql and \pigru, another S-type star. This feature is typical for carbon-rich CSEs and is seen only for S-type AGB stars with a C/O\,$>0.96$ \citep[][their group III stars]{smolders2012_S_SpitzerSurvey}.  MgS is suggested to be present in mantles covering dust grains such as, for example, amorphous alumina. No estimate is available of the amount of S that would be bound in MgS dust. \citet{hony2009_S_ISO} also mentioned the possibility of SiC dust in the spectrum of \waql, at $\sim$11\,\um and $\sim$17\,\um, another dust species typical of C-rich CSEs. It is unclear whether the \waql spectrum shows a contribution from amorphous carbon dust, which typically accounts for several ten percent of the dust mass in carbon-rich CSEs.  Furthermore, the \waql spectrum lacks the amorphous silicate feature at 18\,$\mu$m and lacks a clear silicate feature at 20\,$\mu$m. We are not aware of radiative transfer models of \waql exploring the dust composition in detail. 

\subsection{Influence from the binary companion?}
\citet{danilovich2015_waql_binary} found that the binary companion is a main-sequence dwarf star of spectral type F8\,--\,G0. It is unlikely that this dwarf star has a significant influence on the photochemistry in the CSE of the AGB star. Based on ALMA observations, \cite{ramstedt2017_waql} show that the binary companion has a rather weak, but measurable effect on the CSE morphology. Locally increased densities could cause changes in the circumstellar chemistry. However, the emission lines reported here are detected with APEX, a single-dish telescope, and are most likely more representative of the overall CSE chemistry rather than of localised deviations.

\section{Conclusions}\label{sect:conclusion}
We observed the CSE of the S-type AGB star \waql with APEX at $159-211$\,GHz using the SEPIA/B5 instrument, and in the frequency windows $208.4-239.8$\,GHz, $244.2-255.8$\,GHz, and $260.0-267.8$\,GHz using the PI230 instrument.  We report a number of new detections towards this CSE, including  C$_2$H, SiN, SiC$_2$, and  HC$_3$N, all molecules previously reported only towards C-rich CSEs. We tentatively detect HN$^{13}$C, NaCl and HCO$^+$. We make some first estimates of the molecular abundances, and compare these to abundances reported for CSEs of M-type and C-type AGB stars. We find that the CSE of \waql is more similar to that of a C-type AGB star than to that of an M-type AGB star with similar wind properties, in agreement with the recent chemical equilibrium models of \citet{agundez2020_chemequilibrium}. We cannot currently assess how typical the abundances in \waql's CSE are for S-type CSEs: observations have not yet extensively covered the parameter space of S-type star CSEs, in terms of wind density, $\mbox{C/O}$, and chemical content, and chemical models of the outer CSE of S-type stars are lacking.

Since the ISO observations of \waql show signatures of C$_2$H$_2$ and HCN in the atmosphere and of MgS and SiC in the dust, our detections of molecular gas deemed typical for carbon-rich stars should perhaps not be entirely unexpected. However, previous abundance derivations for commonly targeted molecules such as SiO, SiS, HCN, and CS have consistently led to the conclusion that \waql appears to be a rather typical S-type star, with abundances intermediate to those characteristic for CSEs around C-type and M-type stars. It is only now, with the presented frequency coverage, that the possibly more carbonaceous nature of \waql's molecular CSE becomes apparent.

We hope that our findings on \waql can spark an increased interest in the chemical diversity in the stellar CSEs that comes along with the chemical evolution of the AGB stars themselves. Broader inventories, both in terms of targets and frequency ranges, of the molecular content of CSEs around M-type and S-type stars, in particular, are essential to constrain comprehensive chemical models. Strong constraints on the gas content are also essential to understand the dust formation and dust growth taking place close to the star.

\begin{acknowledgements}
EDB acknowledges financial support from the Swedish National Space Agency. HO acknowledges financial support from the Swedish Research Council. \\
Based on observations with the Atacama Pathfinder EXperiment (APEX) telescope. APEX is a collaboration between the Max Planck Institute for Radio Astronomy, the European Southern Observatory, and the Onsala Space Observatory. 
Swedish observations on APEX are supported through Swedish Research Council grant No 2017-00648.
The APEX observations were obtained under project numbers O-099.F-9306 (SEPIA/B5) and  O-0101.F-9312, O-0102.F-9313, and O-0103.F-9319 (PI230).
\end{acknowledgements}

\addcontentsline{toc}{chapter}{Bibliography}
\bibliographystyle{aa}
\bibliography{38335.bib}

\begin{appendix}

\section{Data presentation}\label{app:data}
\subsection{Data contamination}
We detect five features in our spectra that are leakage from strong lines in the image sideband of a given frequency setup; see Table~\ref{tbl:leakage}. In the SEPIA/B5-data the leakage corresponds to a suppression of the signal at 26\,dB, which is significantly better than the average 18.5\,dB suppression reported for SEPIA/B5 \citep{humphreys2017_sepia,belitsky2018_sepia}. In the PI230 data we find image leakage at suppression levels between 24 and 36\,dB.

\begin{table}[h]
\caption{Image sideband leakage from strong lines. \label{tbl:leakage}}
\begin{tabular}{cccc}
\hline\hline\\[-2ex]
Frequency 	& Molecular	& Fraction of signal 	& Suppres-\\
(GHz)			& 	transition	&	intensity(\%)			& sion (dB)\\
\hline\\[-2ex]
160.3			& SiO\,($4-3$)			& 5.0				& 26\\
209.9			& CO	\,($2-1$)			& 1.5				& 36\\
239.3			& SiO\,($5-4$)			& 1.5				& 36\\
246.1			& HCN\,($3-2$)			& 2.3				& 33\\
267.1			& CS	\,($5-4$)			& 6.2				& 24\\
\hline\hline
\end{tabular}
\end{table}

\subsection{Spectra}
We show the complete spectra obtained with SEPIA/B5 in Fig.~\ref{fig:fullscan_waql_b5} and for PI230 in Fig.~\ref{fig:fullscan_waql_combined}. Both are displayed at a resolution of 7.5\,\kms. 

\subsection{Line properties}
\begin{figure*}
\centering
 \includegraphics[width=0.9\linewidth]{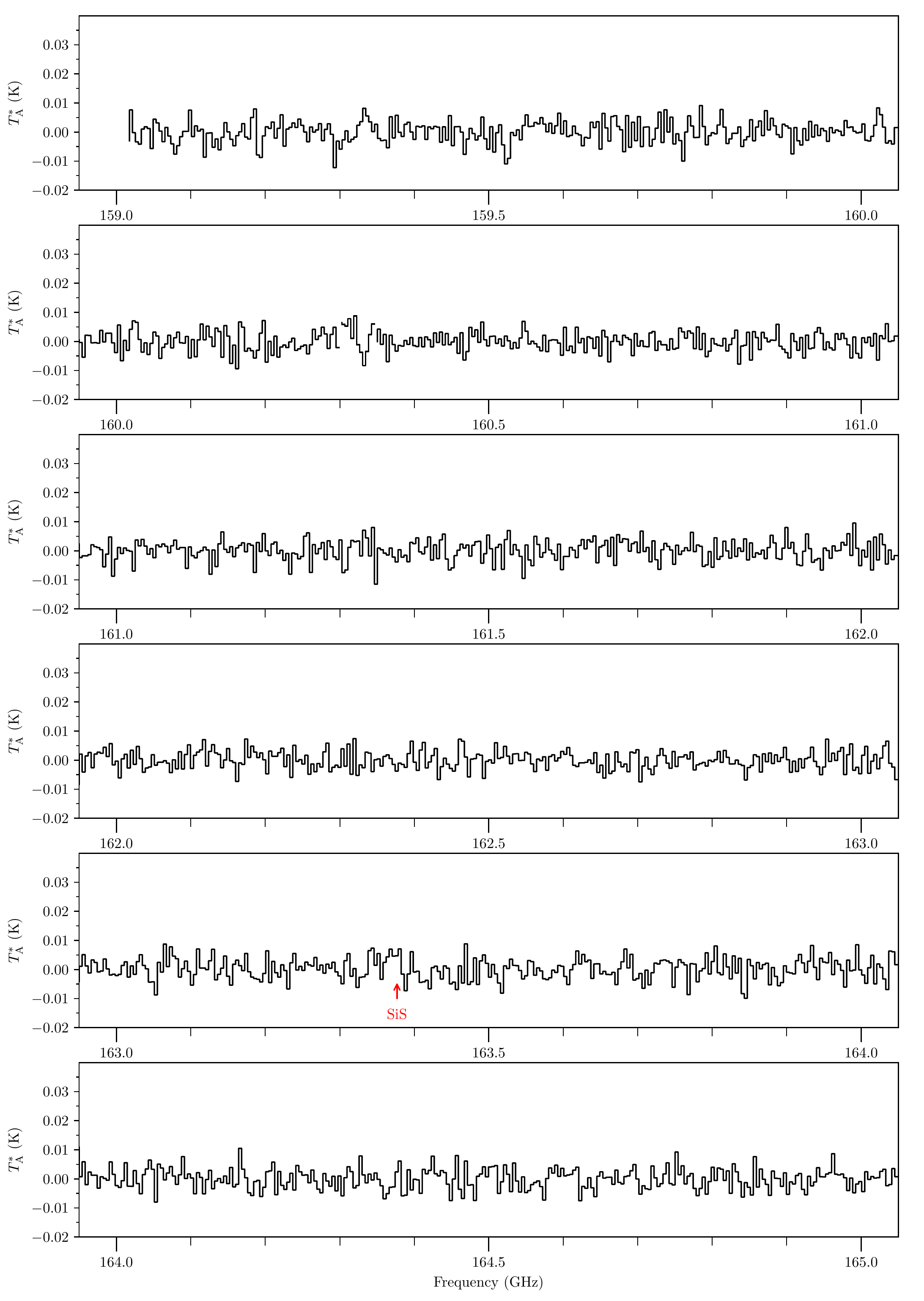}\par 
  \caption{APEX SEPIA/B5 line survey of \waql. Labels show the carrier molecule of the indicated emission. Red labels indicate tentative or unidentified detections. 
  Purple labels (Im(...)) pertain to emission contaminating the signal from the image sideband (see Sect.~\ref{sect:observations}).   For visibility, some parts of the survey were rescaled to fit the vertical scale.  The colour coding of the spectrum corresponds to the following scale factors: \emph{(black)} 1; \emph{(green)} 1/5. 
   Note that the labelling at frequencies above 209.0\,GHz is based on identifications in the PI230 spectrum and is shown here only for reference.  \label{fig:fullscan_waql_b5}}
\end{figure*}

All features that have been detected in our observations and marked in Figs.~\ref{fig:fullscan_waql_b5} and \ref{fig:fullscan_waql_combined} are listed in Table~\ref{tbl:fullscan_waql_combined}. We report the extent of the line in velocity range $[\varv_{\rm min},\varv_{\rm max}]$ and the integrated intensity across this range with a $2\sigma$ error. Comments in the table relate to, for example, tentative detections, blends, or hyperfine structure. Note that the velocity ranges are not necessarily symmetric around 0\,\kms since they were derived to account for weak wings of emission seen for some of the lines. A detailed study of the physical structures within the CSE traced by different molecular transitions is beyond the scope of this paper and would benefit from spatially resolved observations.

\clearpage
\foreach \index in {2, ...,9}
{%
\begin{figure*}
\centering
  \includegraphics[width=0.9\linewidth]{38335_fgA1_\index}\par 
  {\textbf{Fig.~\ref{fig:fullscan_waql_b5}.} Continued.}
\end{figure*}
\clearpage
}

\begin{figure*}
\centering
 \includegraphics[width=0.9\linewidth]{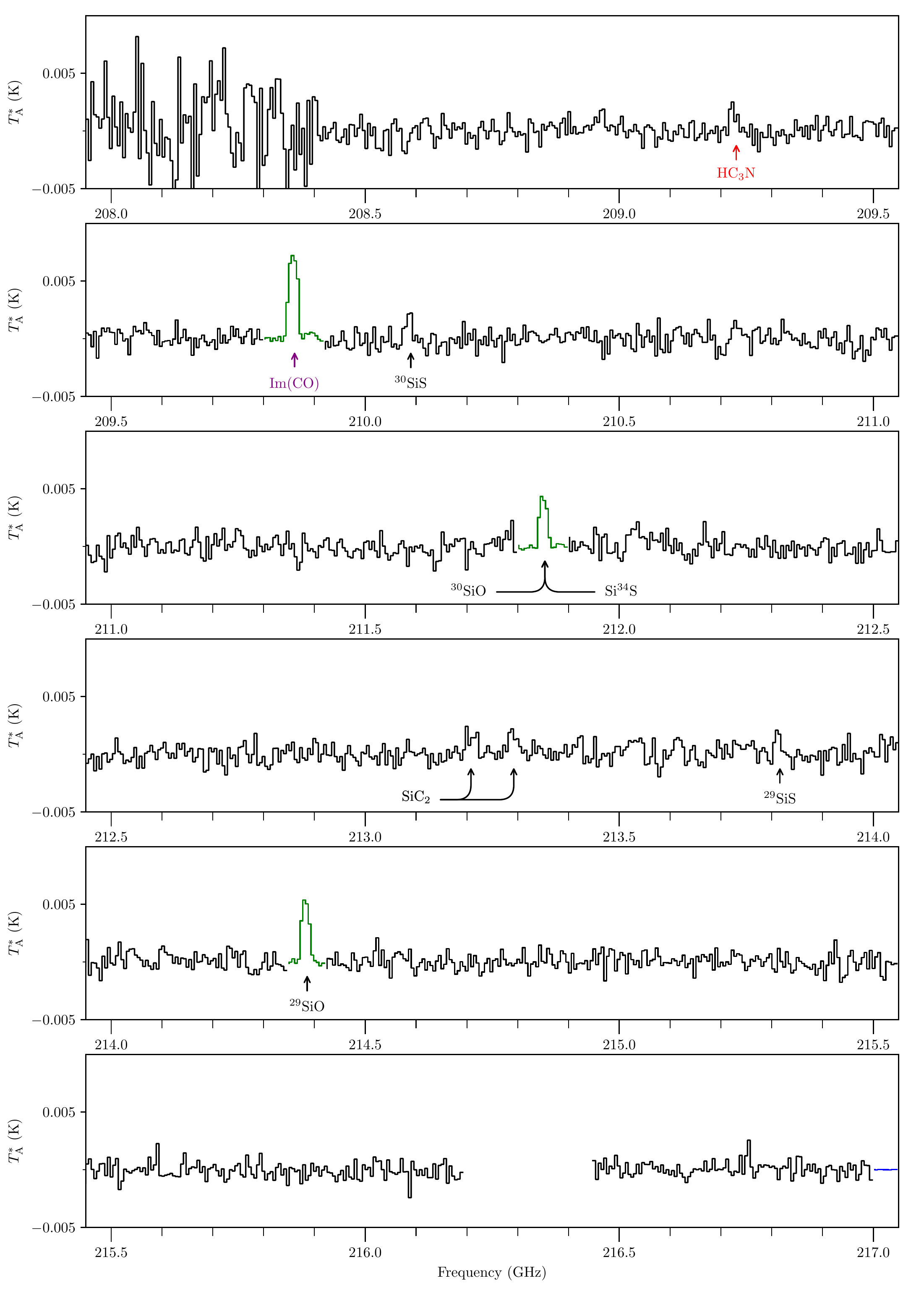}\par 
  \caption{APEX line survey of \waql. Labels show the carrier molecule of the indicated emission. Red labels indicate tentative or unidentified detections. 
  Purple labels (Im(...)) pertain to emission contaminating the signal from the image sideband (see Sect.~\ref{sect:observations}).   For visibility, some parts of the survey were rescaled to fit the vertical scale.  The colour coding of the spectrum corresponds to the following scale factors: \emph{(black)} 1; \emph{(green)} 1/5; \emph{(blue)} 1/25; \emph{(magenta)} 1/125; \emph{(orange)} 1/250. 
   \label{fig:fullscan_waql_combined}}
\end{figure*}
\clearpage
\foreach \index in {2, ...,6} 
{%
\begin{figure*}
\centering
  \includegraphics[width=0.9\linewidth]{38335_fgA2_\index}\par 
  {\textbf{Fig.~\ref{fig:fullscan_waql_combined}.} Continued.}
\end{figure*}
\clearpage
}%

\small
\longtab[2]{
\begin{landscape}
\input{38335_tblA2} 
\end{landscape}
}
\normalsize

\section{Detections}
This appendix features all line emission detected in the spectra in App.~\ref{app:data} other than the lines presented in the body of the paper. In all of these figures the grey-shaded area designates the spectral range specified in Table~\ref{tbl:fullscan_waql_combined} and used to calculate the listed integrated intensities.

\begin{figure*}
\centering
\subfigure[CO]{\includegraphics[width=0.24\linewidth]{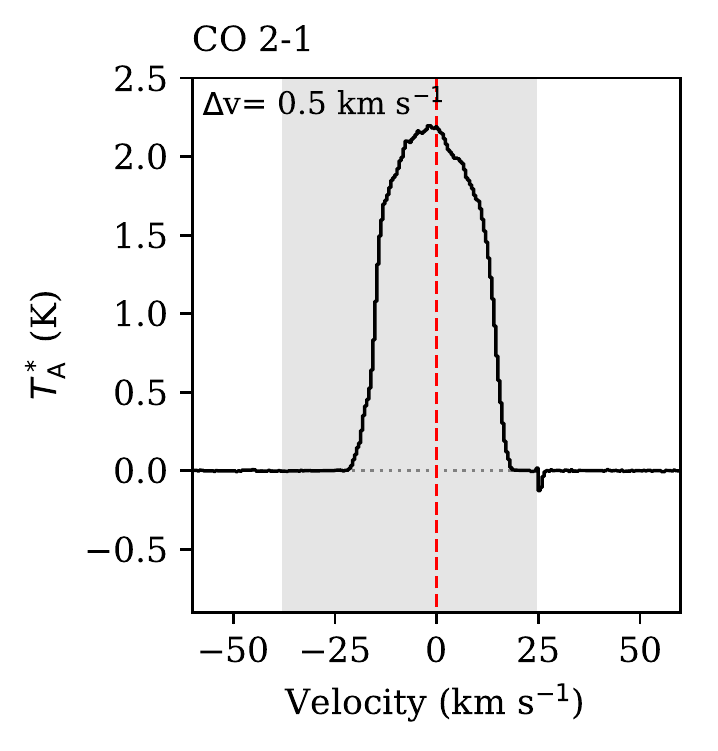}} 
\subfigure[$^{13}$CO]{\includegraphics[width=0.24\linewidth]{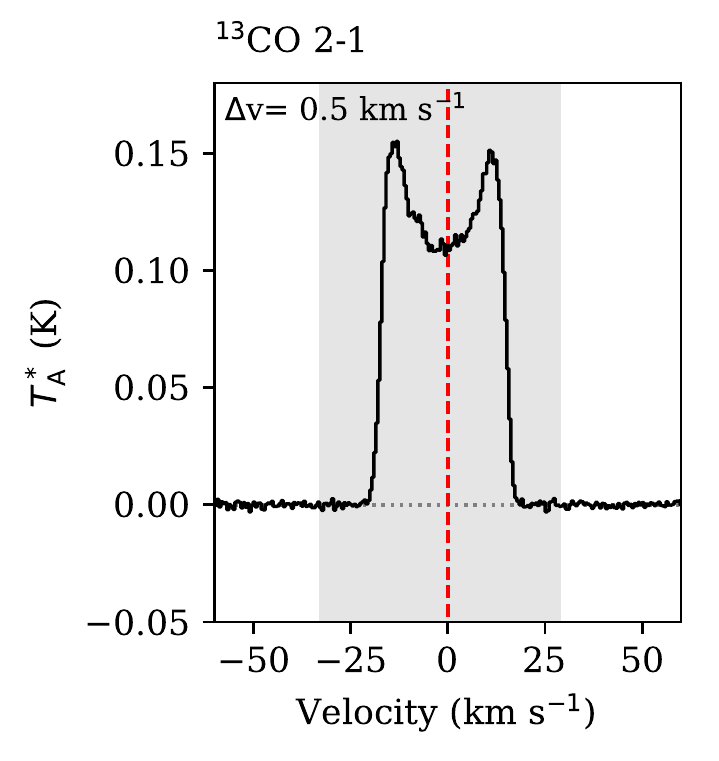}} 
\subfigure[C$^{17}$O]{\includegraphics[width=0.24\linewidth]{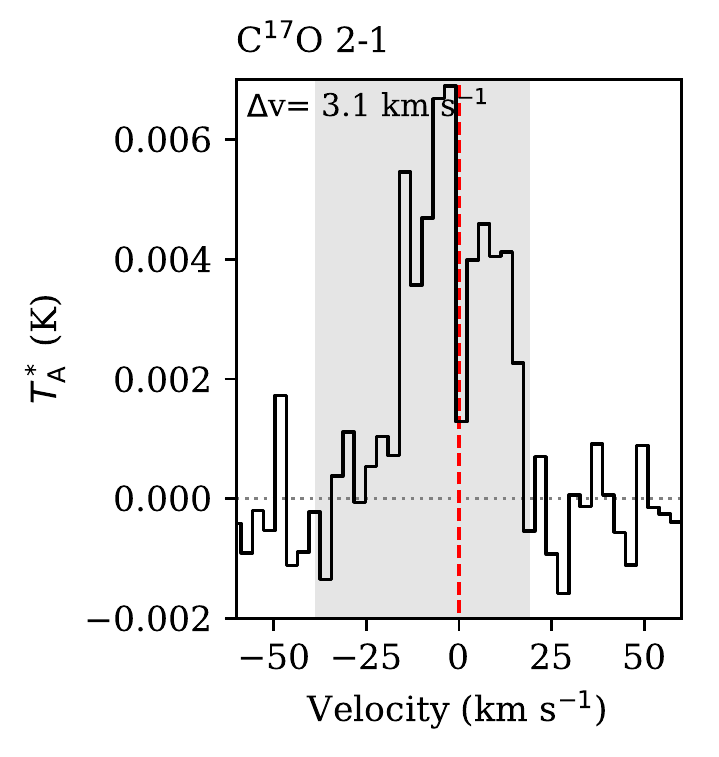}} 
\subfigure[C$^{18}$O]{\includegraphics[width=0.24\linewidth]{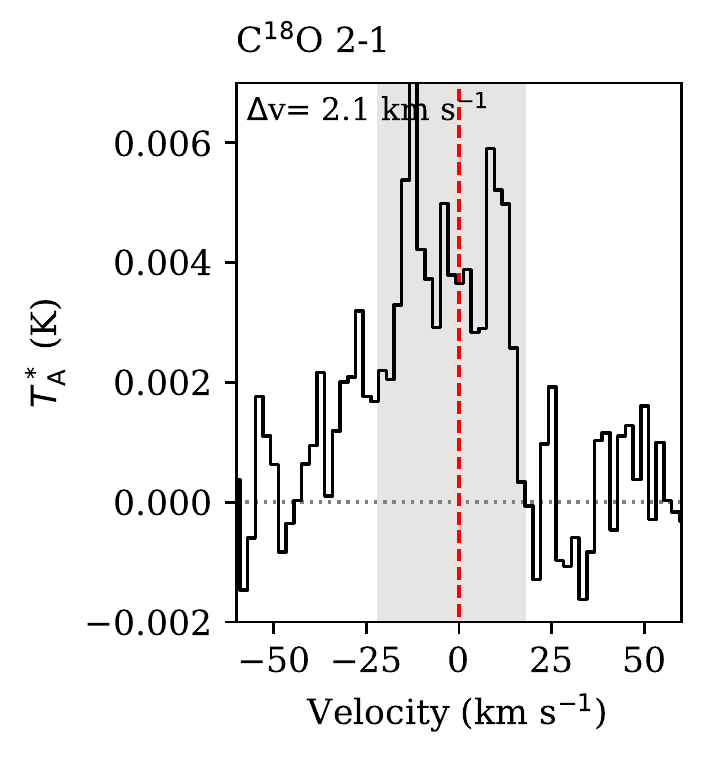}} 
\caption{CO emission.\label{fig:CO}}
\end{figure*}

\begin{figure*}\centering
\subfigure[HCN]{\includegraphics[width=0.48\linewidth]{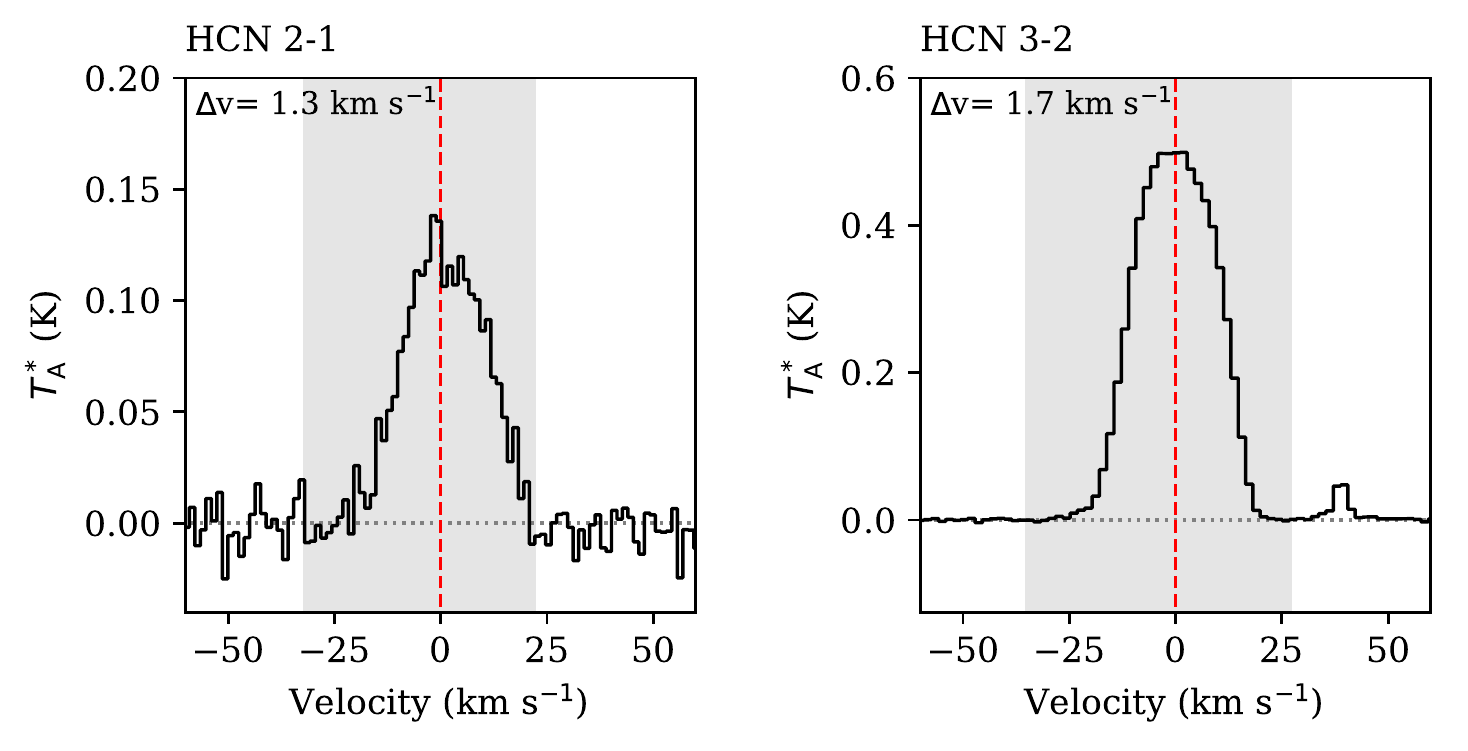}} 
\subfigure[H$^{13}$CN]{\includegraphics[width=0.24\linewidth]{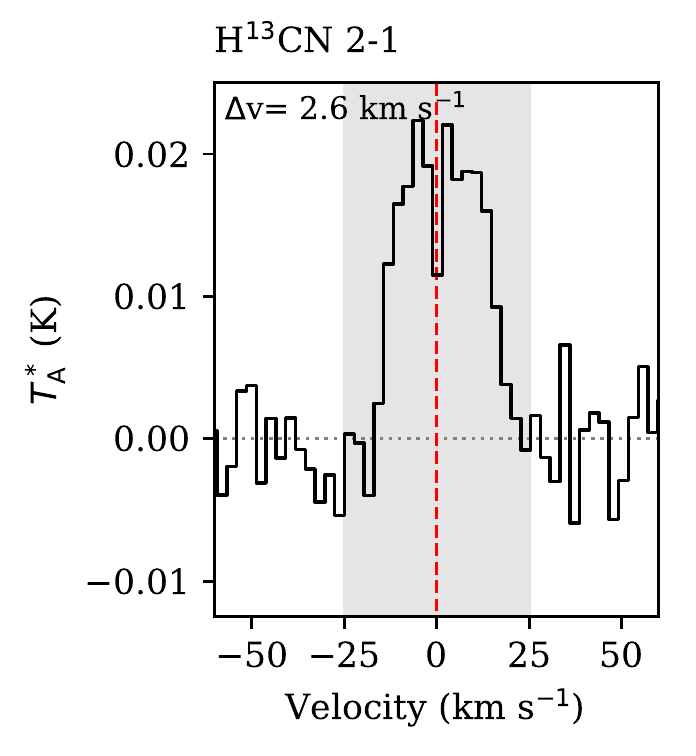}} 
\subfigure[HCN, $v_2=1$]{\includegraphics[width=0.48\linewidth]{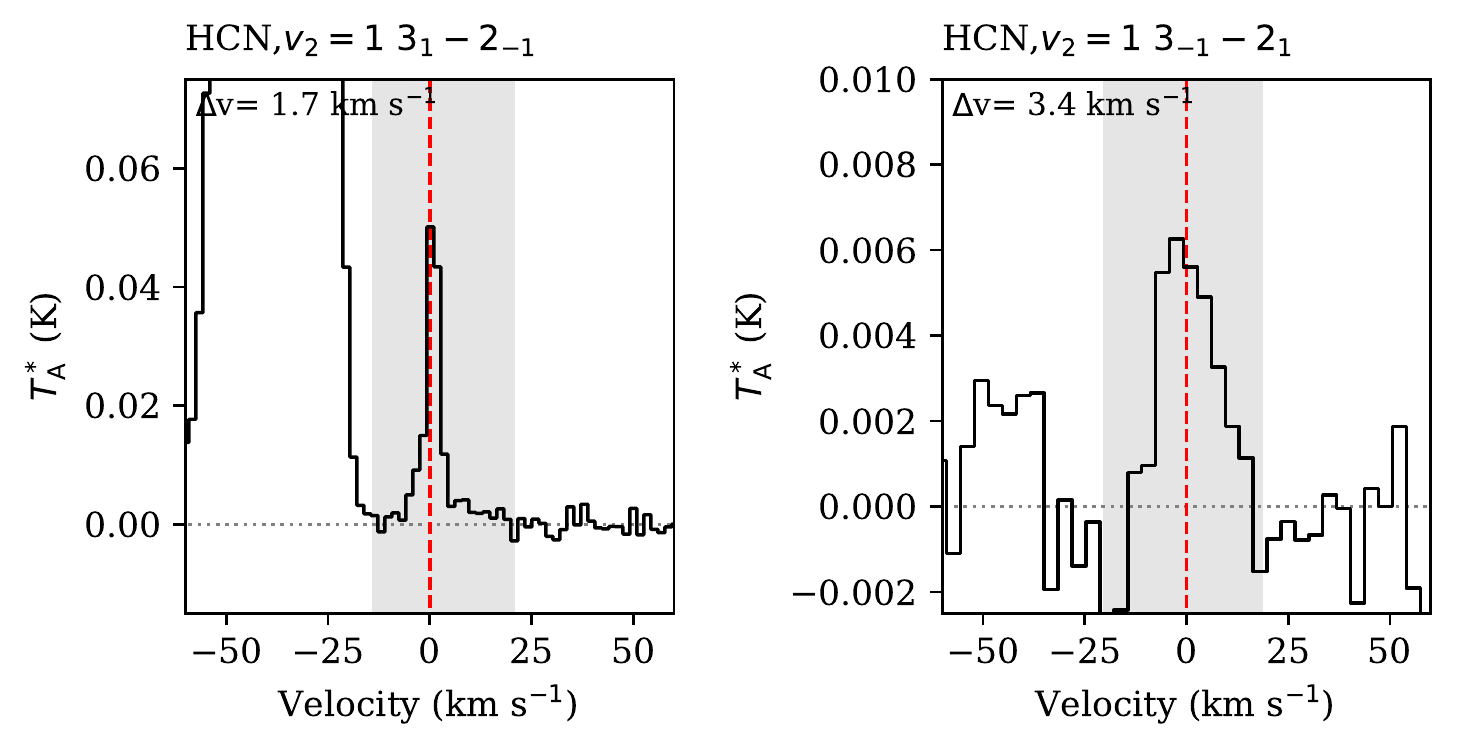}} 
\caption{HCN emission.\label{fig:HCN}}
\end{figure*}

\begin{figure*}\centering
\subfigure[SiO]{\includegraphics[width=0.72\linewidth]{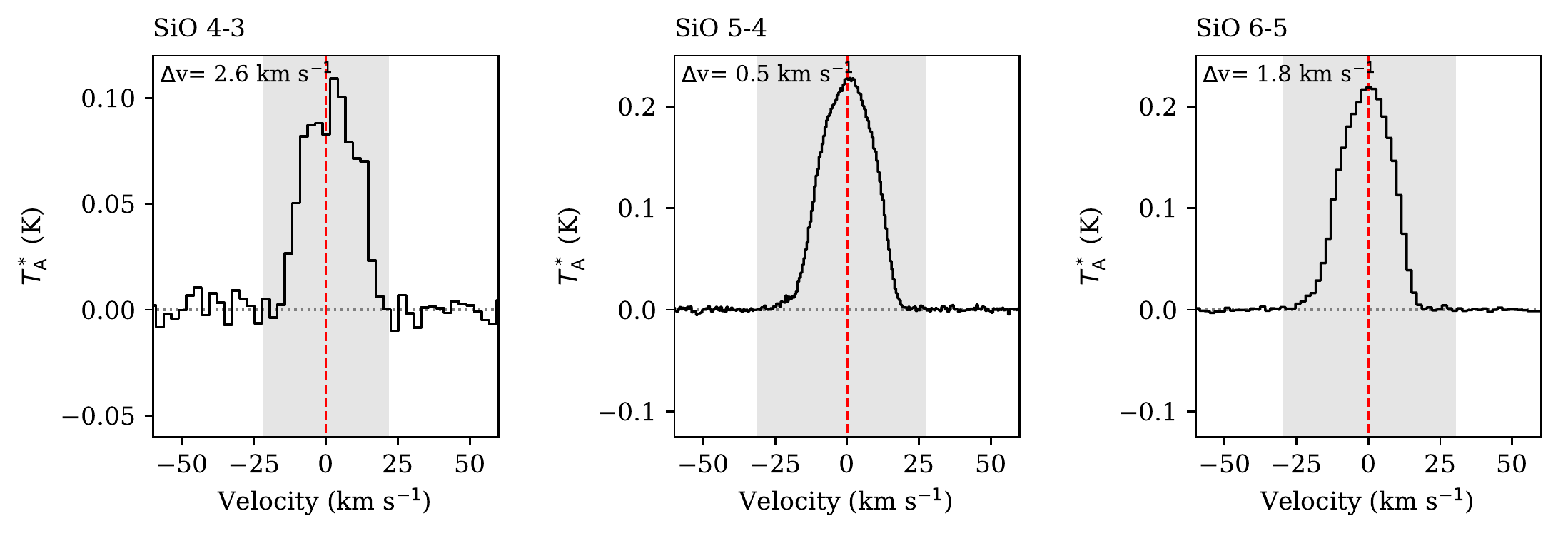}} 
\subfigure[$^{29}$SiO]{\includegraphics[width=0.48\linewidth]{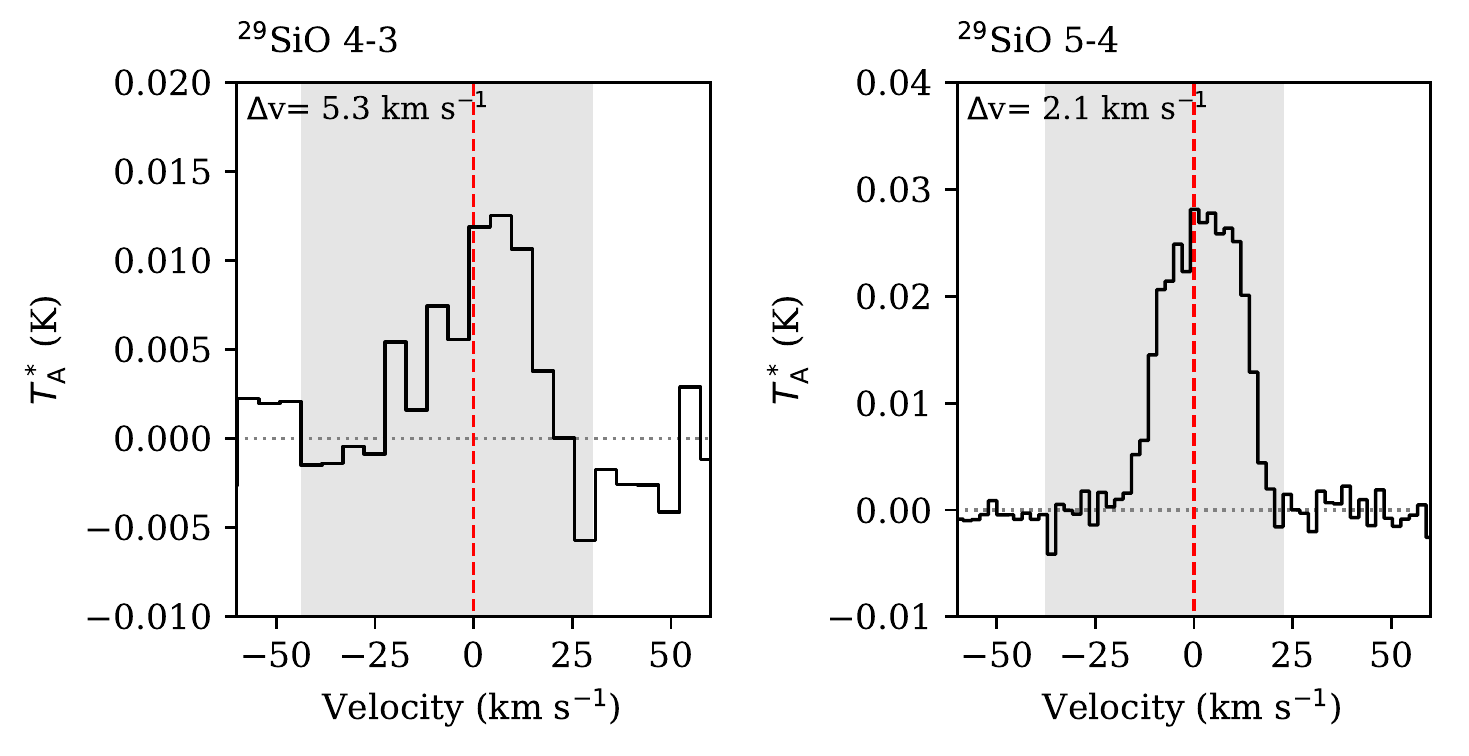}} 
\subfigure[Si$^{17}$O]{\includegraphics[width=0.24\linewidth]{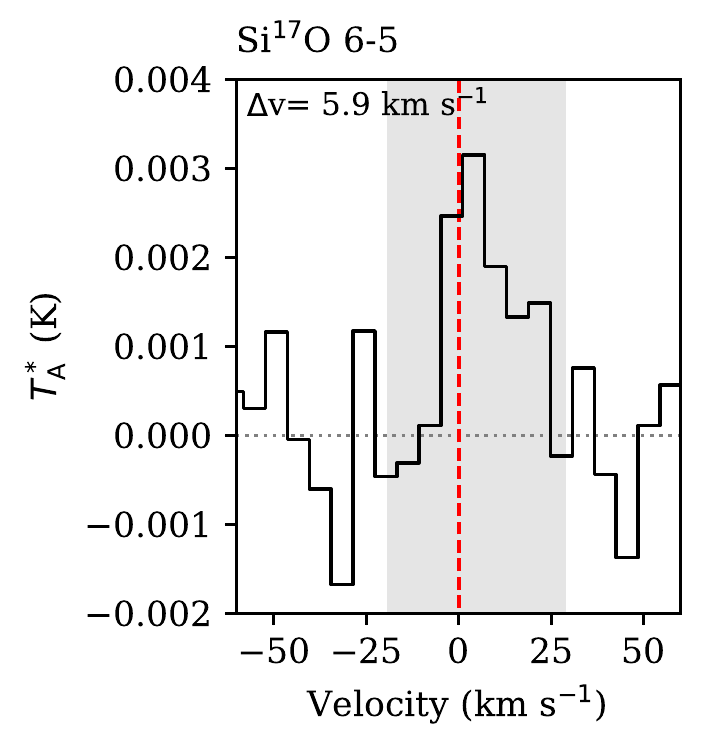}} 
\subfigure[$^{30}$SiO. Note that the $J=5-4$ emission is blended with emission from Si$^{34}$S\,($J=12-11$).]{\includegraphics[width=0.72\linewidth]{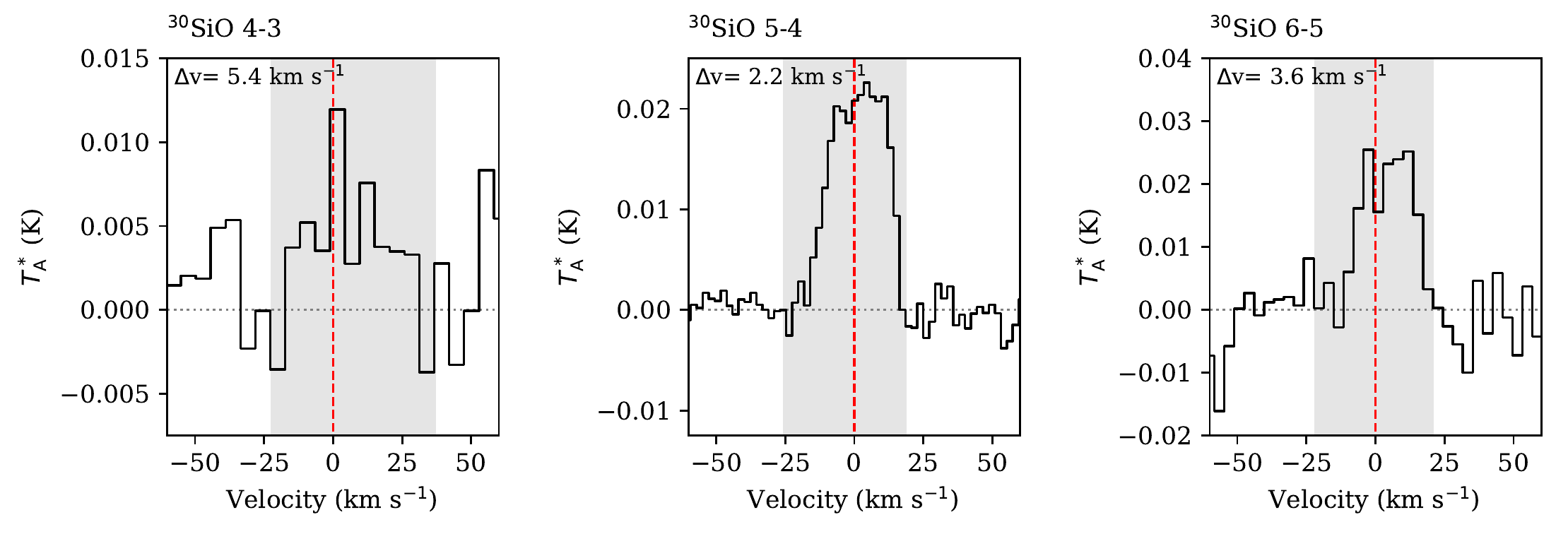}} 
\caption{SiO emission.\label{fig:SiO}}
\end{figure*}

\begin{figure*}\centering
\subfigure[SiS]{\includegraphics[width=\linewidth]{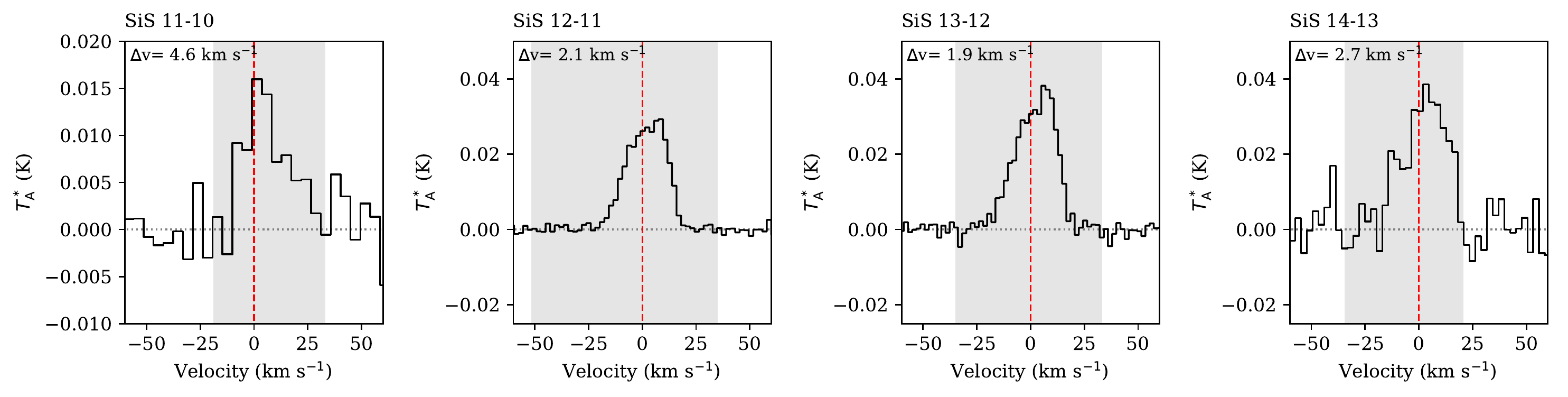}} 
\subfigure[$^{29}$SiS]{\includegraphics[width=0.72\linewidth]{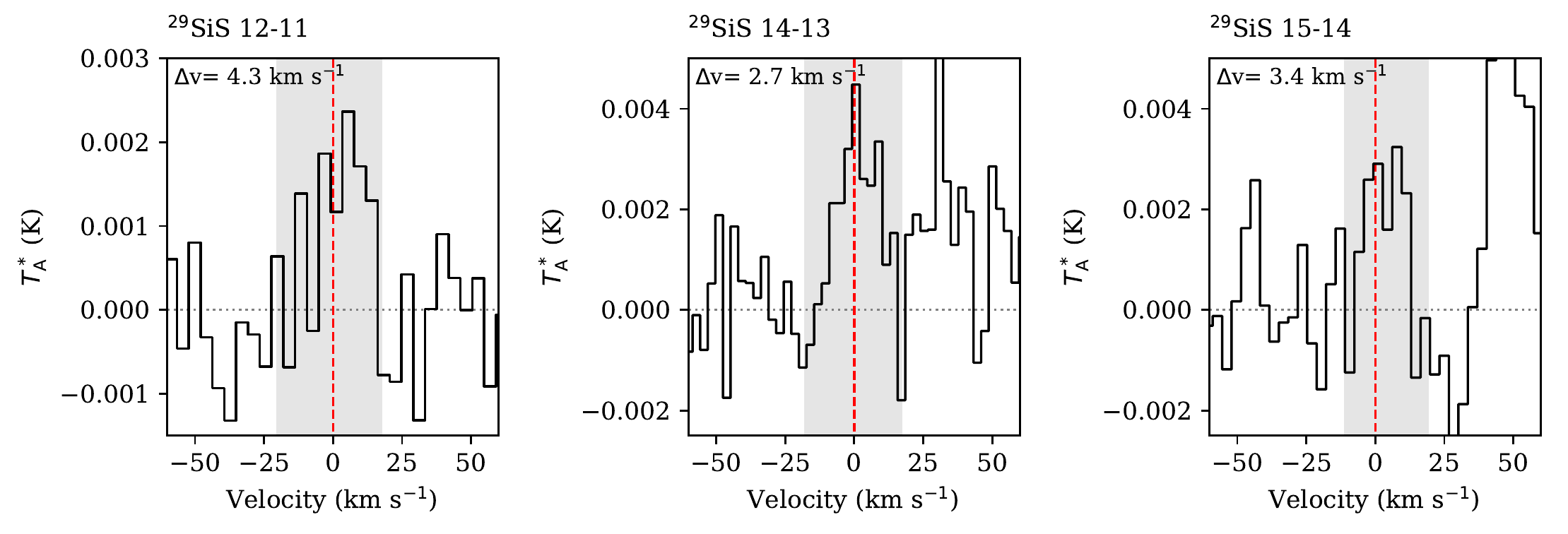}}\\
\subfigure[$^{30}$SiS]{\includegraphics[width=0.96\linewidth]{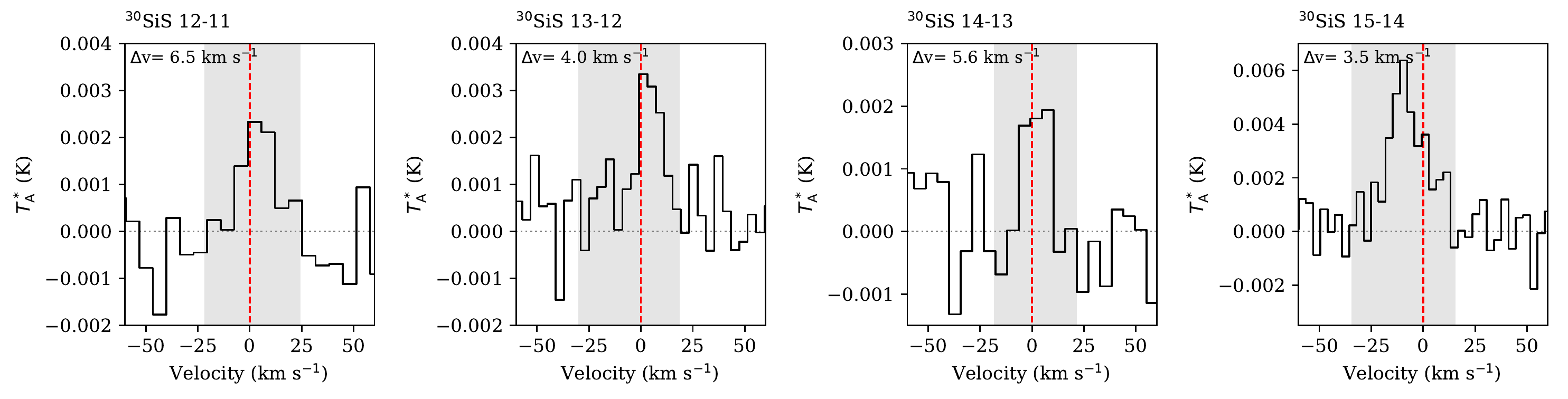}}\\
\subfigure[Si$^{34}$S. Note that the $J=12-11$ emission is blended with emission from $^{30}$SiO\,($J=5-4$)]{\includegraphics[width=0.72\linewidth]{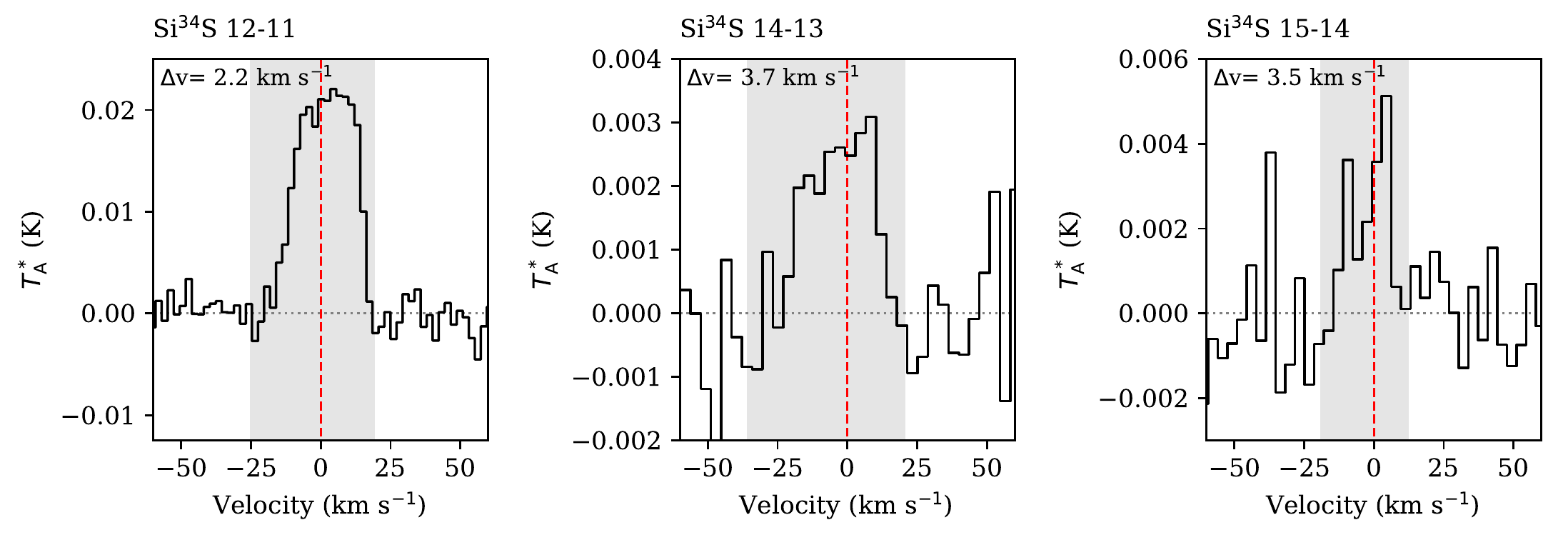}}
\caption{SiS emission.\label{fig:SiS}}
\end{figure*}

\begin{figure*}
\centering
\subfigure[CS]{\includegraphics[width=0.24\linewidth]{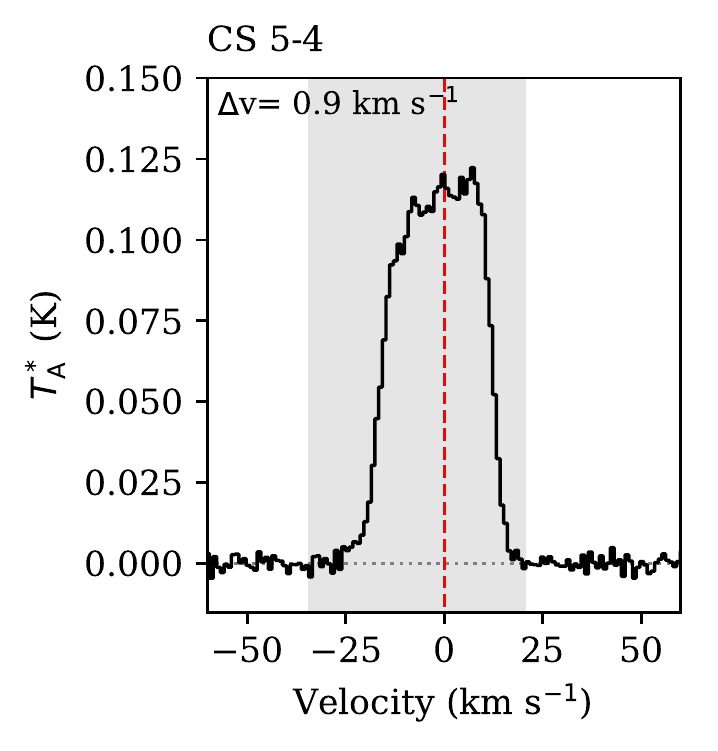}}
\subfigure[C$^{34}$S]{\includegraphics[width=0.24\linewidth]{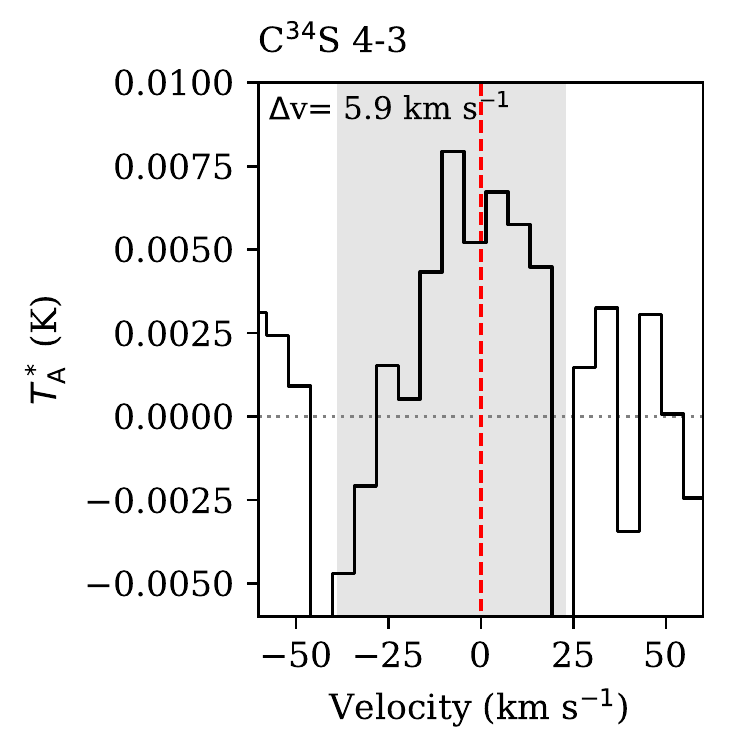}}
\subfigure[$^{13}$CS]{\includegraphics[width=0.24\linewidth]{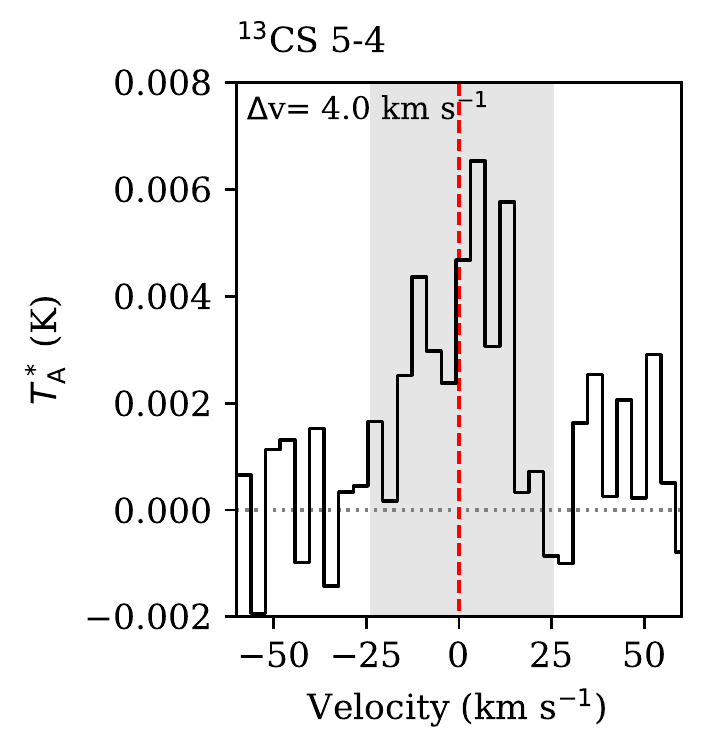}}
\caption{CS emission. \label{fig:CS}}
\end{figure*}

\begin{figure*}\centering
\includegraphics[width=\linewidth]{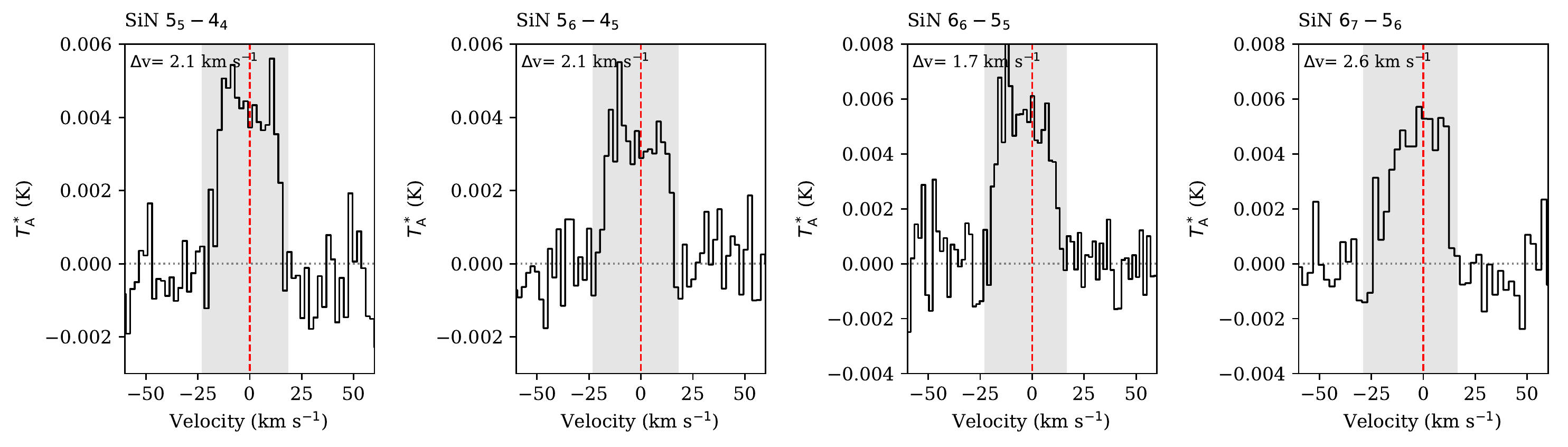}
\caption{SiN emission. Rest frequencies were calculated by weighting the hyperfine component frequencies with the intrinsic component intensities.\label{fig:SiN}}
\end{figure*}

\begin{figure*}\centering
\includegraphics[width=.8\linewidth]{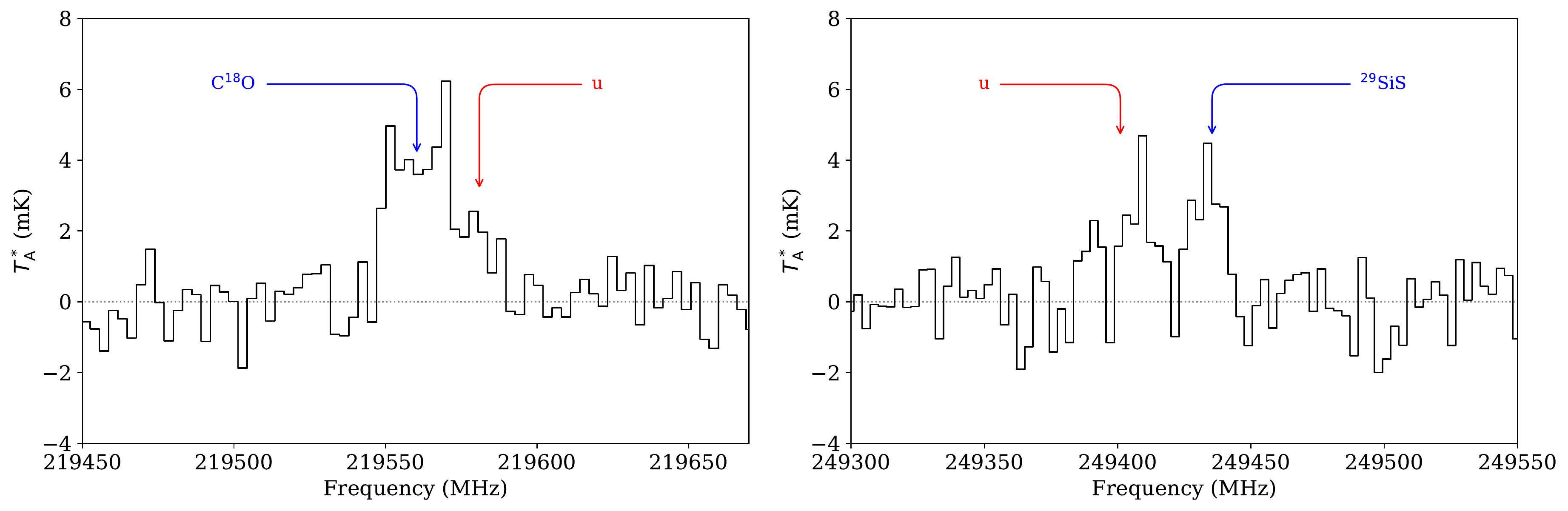}
\caption{Unidentified emission features. \label{fig:ulines}}
\end{figure*}

\clearpage

\section{Fractional abundances for M-, S-, C-type CSEs}
The peak fractional abundances used to compile Fig.~\ref{fig:compabundance} are listed in Table~\ref{tbl:compabundance}, along with the references to the relevant studies and a marker in case of upper limits. We refer to those papers for  presentation of and discussion on the modelling procedures used to estimate the abundances and the target samples used to obtain median abundance values for M-type, S-type, and C-type stars.

\def\tblfootfile{Mytblfoot.tex}
\newwrite\file
\immediate\openout\file=\tblfootfile
\newcounter{tblciteno}
\newcommand{\tblfootcitation}[1]{\tablefoottext{\ref{tblcite:#1}}{\citet{#1}}}	
\newcommand{\tblcitemark}[1]{\tablefootmark{\ref{tblcite:#1}}}					
\newcommand{\buildtblfoot}[1]{\immediate\write\file{\unexpanded{\tblfootcitation{#1}}}}
\newcommand{\tblcite}[1]{\refstepcounter{tblciteno}\label{tblcite:#1} \tblcitemark{#1} \buildtblfoot{#1}}		
\begin{table*}
\centering
\caption{Peak fractional abundances used in Fig.~\ref{fig:compabundance}. \label{tbl:compabundance}}
\begin{tabular}{p{1.2cm} | p{1.25cm}p{1.25cm}p{1.25cm}  | p{1.25cm}p{2.4cm}p{1.25cm}  | p{1.25cm}p{1.7cm}p{1.25cm} }
\hline\hline\\[-2ex]		
				
Molecule& \rdor	& \iktau	& M-type& \chicyg	&\waql	& S-type& \vaql	& \irc	& C-type\\
\hline \\[-2ex]

HCN
& $9.0(-8)$\tblcite{schoeier2013_hcn} & $2.2(-7)$\tblcite{decin2010_nlte}& $1.2(-7)$\tblcitemark{schoeier2013_hcn}
& $5.0(-6)$\tblcite{schoeier2011_chicyg}& $3.1\pm0.1(-6)$\tblcite{danilovich2014}& $7.0(-7)$\tblcitemark{schoeier2013_hcn}
& $1.3(-5)$\tblcitemark{schoeier2013_hcn}& $2.0(-5)$\tblcite{agundez2012_innerlayers_irc10216}& $2.5(-5)$\tblcitemark{schoeier2013_hcn}\\

CS
& -- & $4.0(-8)$ \tblcitemark{decin2010_nlte}& $8.0(-8)$\tblcite{danilovich2018_cs_sis}
& $1.0(-7)$  \tblcitemark{schoeier2011_chicyg}& $1.2\pm0.05(-6)$\tblcite{brunner2018_waql_alma}& $5.0(-7)$ \tblcitemark{danilovich2018_cs_sis}
& $1.0(-5)$\tblcite{massalkhi2019_cs_sio_sis}& $4.0(-6)$\tblcitemark{agundez2012_innerlayers_irc10216}& $7.0(-6)$ \tblcitemark{massalkhi2019_cs_sio_sis}\\

SiS
& --  & $5.5(-6)$ \tblcitemark{decin2010_nlte}& $1.1(-6)$ \tblcitemark{danilovich2018_cs_sis}
& $1.8(7-)$ \tblcitemark{schoeier2011_chicyg}& $1.5\pm0.05(-6)$\tblcitemark{brunner2018_waql_alma}& $6.0(-7)$ \tblcitemark{danilovich2018_cs_sis}
& $2.3(-6)$\tblcitemark{massalkhi2019_cs_sio_sis}\tablefootmark{$\dagger$}& $3.0(-6)$\tblcitemark{agundez2012_innerlayers_irc10216}& $1.0(-5)$  \tblcitemark{massalkhi2019_cs_sio_sis}\\

SiO
& $2.0(-5)$\tblcite{debeck2018_rdor}& $2.0(-7)$  \tblcitemark{decin2010_nlte}& $5.0(-6)$\tblcite{gonzalezdelgado2003}
& $1.3(-5)$  \tblcitemark{schoeier2011_chicyg}& $2.9\pm0.7(-6)$\tblcitemark{danilovich2014}& $6.0(-6)$\tblcite{ramstedt2009_sio}
& $2.5(-6)$\tblcitemark{massalkhi2019_cs_sio_sis}& $1.8(-7)$\tblcitemark{agundez2012_innerlayers_irc10216}& $3.8(-6)$\tblcite{schoeier2006_sio}\\

H$_2$O
& $2.5(-4)$\tblcite{maercker2016_water} & $4.2(-4)$ \tblcitemark{maercker2016_water}& --  
& $1.2(-5)$  \tblcitemark{schoeier2011_chicyg}& $1.5\pm0.5(-5)$\tblcitemark{danilovich2014}& --
& -- & $1.0(-7)$\tblcitemark{agundez2012_innerlayers_irc10216}& --  \\

NH$_3$
& -- & $6.0(-7)$\tblcite{wong2018_nh3}& --
& $1.8(-8)$ \tblcitemark{schoeier2011_chicyg}\tablefootmark{$\dagger$}& $1.7\pm1.0(-5)$\tblcitemark{danilovich2014}& --
& --& $6.0(-8)$\tblcite{schmidt2016_irc10216_nh3}& --\\

CN
& -- & $3.0(-8)$ \tblcitemark{decin2010_nlte} & --
& $1.9(-5)$ \tblcitemark{schoeier2011_chicyg}& $5.7(-6)$\tblcitemark{danilovich2014}& --
& --& $1.7(-6)$\tblcite{agundez2009_phd}& --\\

SiC$_2$
& -- & --& --
& --& $5.0\pm2.0(-7)$& --
& $2.0(-5)$\tblcite{massalkhi2018_sic2}& $3.7(-7)$\tblcitemark{massalkhi2018_sic2}& $6.3(-6)$\tblcitemark{massalkhi2018_sic2}\\

C$_2$H
& -- & --& --
& --& $1.0(-5)$\tablefootmark{$\ddagger$}& --
& --& $2.0(-5)$\tblcite{debeck2012_cch}& --\\

HNC
& -- & $8.0(-9)$\tblcite{velillaprieto2017_iktau_iram}& --
& --& $3.1(-8)$ \tablefootmark{$\ddagger$}\tablefootmark{$*$}& --
& --& $7.0(-8)$\tblcite{daniel2012_hnc}& --\\

HC$_3$N
& -- & --& --
& --& $8.4(-7)$ \tablefootmark{$\dagger$}& --
& --& $1.4(-6)$\tblcitemark{agundez2009_phd}& --\\

SiN
& -- & --& --
& --& $4.0(-8)$\tablefootmark{$\ddagger$}& --
& --& $8.0(-9)$\tblcitemark{agundez2009_phd}& --\\

H$_2$S
& -- & $1.6(-6)$\tblcite{danilovich2017_h2s}& --
& --& $5.0(-7)$\tablefootmark{$\dagger$}& --
& --& $4.0(-9)$\tblcitemark{agundez2012_innerlayers_irc10216}& --\\

PN
& -- & $3.0(-7)$\tblcite{debeck2013_popn}& --
& --& $3.0(-7)$\tablefootmark{$\dagger$}& --
& --& $1.0(-9)$\tblcitemark{agundez2009_phd}& --\\

HCO$^+$
& -- & $1.0(-8)$ \tblcitemark{velillaprieto2017_iktau_iram}& --
& --& $7.0(-8)$ \tablefootmark{$\dagger$}& --
& --& $4.1(-9)$\tblcite{pulliam2011_hcoplus}& --\\

NaCl
& -- & $3.1(-7)$ \tblcitemark{velillaprieto2017_iktau_iram}& --
& --& $3.1(-8)$ \tablefootmark{$\dagger$}& -- 
& --& $1.8(-9)$\tblcitemark{agundez2012_innerlayers_irc10216}& --\\

SO
& $6.7(-6)$\tblcite{danilovich2016_sulphur} & $1.0(-6)$  \tblcitemark{danilovich2016_sulphur}& --
& --& $1.0(-6)$\tablefootmark{$\dagger$}& --
& --& --& --\\

SO$_2$
& $5.0(-6)$  \tblcitemark{danilovich2016_sulphur} & $8.6(-7)$ \tblcitemark{danilovich2016_sulphur}& --
& --& $5.0(-6)$\tablefootmark{$\dagger$}& --
& --& --& --\\
\hline
\end{tabular}
\immediate\closeout\file

\normalsize
\tablefoot{
Fractional abundances  $a\times10^b$ are listed as $a(b)$,  $(a\pm e)\times10^b$  as $a\pm e (b)$. Where no reference is given, the abundance is estimated in this work.
\tablefoottext{$\dagger$}{Upper limit.}
\tablefoottext{$\ddagger$}{We propose an order of magnitude uncertainty, based on the uncertainties described in our analysis.}
\tablefoottext{$*$}{The HNC abundance was derived assuming $\mbox{HNC/H}_2=0.01 \times \mbox{HCN/H}_2$.} 
 }
\tablebib{
\input{\tblfootfile}
}

\end{table*}
\normalsize

 \end{appendix}

\end{document}

%% file: 38335_tblA2.tex
\begin{longtable}{rlcrccccrp{5cm}} 
\caption{Molecular lines detected in our survey towards \waql.  \label{tbl:fullscan_waql_combined}}\\
\hline\hline\\[-2ex]
No. & Molecule & Transition & $E_{\mathrm{u}}/k$ & Freq. & Res. & $v_{\mathrm{min}}$ & $v_{\mathrm{max}}$ & $I_{\mbox{int}}$ & Comments \\

 &  &  & (K) & (MHz) & (MHz) & (km\,s$^{-1}$) & (km\,s$^{-1}$) & (mK\,km\,s$^{-1}$) &  \\

\hline\\[-2ex]
\endfirsthead
\multicolumn{3}{l}{\textbf{{\tablename} \thetable{}.} Continued} \\[0.5ex]
\hline\hline\\[-2ex]
No. & Molecule & Transition & $E_{\mathrm{u}}/k$ & Freq. & Res. & $v_{\mathrm{min}}$ & $v_{\mathrm{max}}$ & $I_{\mbox{int}}$ & Comments \\

 &  &  & (K) & (MHz) & (MHz) & (km\,s$^{-1}$) & (km\,s$^{-1}$) & (mK\,km\,s$^{-1}$) &  \\

\hline\\[-2ex]
\endhead
\hline\\[-2ex]
\multicolumn{3}{l}{Continued on next page\ldots} \\[0.5ex]
\endfoot
\\[-1.8ex]
\endlastfoot
\hline \\[-2ex]
1 & Im(SiO) & $/$ & -- & 160292.000 & -- & -- & -- & -- & Image contamination (Table~\ref{tbl:leakage}). \\
2 & SiS (t) & $9-8$ & 39.2 & 163376.781 & 3.1 & -25.9 & 31.3 & 192.2$\pm$  15.4 & Tentative. \\
3 & $^{30}$SiO (t) & $4-3$ & 20.3 & 169486.875 & 3.1 & -17.2 & 32.0 & 254.8$\pm$  25.4 & Tentative. \\
4 & $^{29}$SiO & $4-3$ & 20.6 & 171512.797 & 3.1 & -38.5 & 25.0 & 295.2$\pm$  45.8 &  \\
5 & H$^{13}$CN & $2-1$ & 12.4 & 172677.844 & 1.5 & -22.7 & 22.8 & 543.9$\pm$  75.6 &  \\
6 & HC$_3$N (t) & $19-18$ & 83.0 & 172849.297 & 2.3 & -16.7 & 7.5 & 72.6$\pm$  11.9 & Tentative. \\
7 & SiO & $4-3$ & 20.8 & 173688.234 & 1.5 & -19.3 & 19.2 & 2313.0$\pm$ 179.0 &  \\
8 & C$_2$H (t) & $N=2-1, J=5/2-3/2$ & 12.6 & 174664.000 & 6.1 & -65.9 & 61.1 & 141.0$\pm$  61.1 & Hyperfine structure. Tentative. \\ 
9 & C$_2$H (t) & $N=2-1, J=3/2-1/2$ & 12.6 & 174725.000 & 6.1 & -54.6 & 38.8 & 68.9$\pm$  35.2 & Hyperfine structure. Tentative. \\
10 & HCN & $2-1$ & 12.8 & 177261.109 & 0.8 & -31.1 & 21.2 & 3005.2$\pm$ 321.9 &  \\
11 & HCO$^+$ (t) & $2-1$ & 12.8 & 178375.062 & 2.8 & -26.1 & 26.7 & 47.5$\pm$  11.6 & Tentative. \\
12 & C$^{34}$S (t) & $4-3$ & 23.1 & 192818.453 & 3.8 & -18.1 & 17.0 & 163.8$\pm$  30.4 & Tentative. \\
13 & SiS & $11-10$ & 57.5 & 199672.234 & 3.1 & -14.4 & 28.6 & 344.1$\pm$  48.0 & Asymmetric. \\
14 & HC$_3$N (t) & $23-22$ & 120.5 & 209230.234 & 3.7 & -26.3 & 30.6 & 49.3$\pm$   5.9 & Tentative. \\
15 & Im(CO) & $/$ & -- & 209861.000 & -- & -- & -- & -- & Image contamination (Table~\ref{tbl:leakage}). \\
16 & $^{30}$SiS & $12-11$ & 65.5 & 210089.625 & 4.5 & -15.2 & 17.9 & 41.9$\pm$   4.2 &  \\
17 & $^{30}$SiO & $5-4$ & 30.5 & 211853.469 & 1.5 & -23.7 & 16.8 & 561.1$\pm$  60.5 & Blend with Si$^{34}$S. \\
18 & Si$^{34}$S & $12-11$ & 66.1 & 211853.734 & 1.5 & -23.3 & 17.2 & 557.1$\pm$  64.1 & Blend with $^{30}$SiO. \\
19 & SiC$_2$ & $9_{4,6}-8_{4,5}$ & 82.3 & 213208.031 & 4.6 & -26.4 & 19.0 & 62.5$\pm$   9.4 &  \\
20 & SiC$_2$ & $9_{4,5}-8_{4,4}$ & 82.3 & 213292.344 & 4.6 & -36.3 & 29.2 & 65.1$\pm$   5.4 &  \\
21 & $^{29}$SiS & $12-11$ & 66.7 & 213816.141 & 3.1 & -16.3 & 13.6 & 31.3$\pm$  10.1 &  \\
22 & $^{29}$SiO & $5-4$ & 30.9 & 214385.750 & 1.5 & -35.7 & 20.8 & 680.1$\pm$  49.7 &  \\
23 & SiO & $5-4$ & 31.3 & 217104.922 & 0.4 & -31.0 & 26.9 & 5226.6$\pm$ 124.5 &  \\
24 & $^{13}$CN & $N=2-1, J=3/2-1/2$ & 15.7 & 217280.000 & 1.5 & -71.5 & 28.3 & 283.2$\pm$63.3 & Hyperfine structure. \\
25 & $^{13}$CN & $N=2-1, J=5/2-3/2$ & 15.7 & 217405.000 & 1.8 & -64.1 & -10.1 & 101.2$\pm$  35.4 & Hyperfine structure. \\
26 & $^{13}$CN & $N=2-1, J=5/2-3/2$ & 15.7 & 217469.000 & 1.8 & -42.0 & 17.6 & 126.6$\pm$ 39.3 & Hyperfine structure. \\
27 & SiS & $12-11$ & 68.0 & 217817.656 & 1.5 & -49.6 & 32.8 & 674.9$\pm$  54.8 &  \\
28 & SiN & $N=5-4, J=9/2-7/2$ & 31.4 & 218008.000 & 1.5 & -20.5 & 16.9 & 134.9$\pm$  35.5 &  \\
29 & HC$_3$N (t) & $24-23$ & 131.0 & 218324.719 & 1.8 & -17.7 & 18.3 & 62.4$\pm$  12.6 & Tentative. \\
30 & SiN & $N=5-4, J=11/2-9/2$ & 31.5 & 218513.000 & 1.5 & -20.7 & 16.5 & 112.6$\pm$  29.7 &  \\
31 & C$^{18}$O & $2-1$ & 15.8 & 219560.359 & 1.5 & -20.1 & 15.9 & 143.1$\pm$  44.9 & Blend with unidentified feature. \\
32 & u & $/$ & -- & 219581.000 & 1.8 & -18.2 & 43.6 & 186.3$\pm$  89.5 & No candidate. Blend with C$^{18}$O. \\
33 & $^{13}$CO & $2-1$ & 15.9 & 220398.688 & 0.4 & -32.5 & 28.7 & 4182.7$\pm$114.8 &  \\
34 & SiC$_2$ & $10_{0,10}-9_{0,9}$ & 59.8 & 220773.688 & 1.5 & -31.4 & 17.4 & 106.8$\pm$  26.1 &  \\
35 & SiC$_2$ & $9_{2,7}-8_{2,6}$ & 60.2 & 222009.391 & 1.5 & -25.7 & 14.6 & 106.6$\pm$  28.1 &  \\
36 & C$^{17}$O & $2-1$ & 16.2 & 224714.203 & 2.3 & -35.9 & 16.2 & 154.6$\pm$  31.6 &  \\
37 & CN & $N=2-1, J=3/2-3/2$ & 16.3 & 226310.000 & 1.5 & -89.8 & 46.9 &  1145.8$\pm$ 208.0 & Hyperfine structure. \\
38 & CN & $N=2-1, J=3/2-1/2$ & 16.3 & 226659.000 & 1.5 & -54.0 & 77.8 &  4099.5$\pm$ 182.0 & Hyperfine structure. \\
39 & CN & $N=2-1, J=5/2-3/2$ & 16.3 & 226876.000 & 1.5 & -45.1 & 33.6 &  3513.1$\pm$ 117.0 & Hyperfine structure. \\
40 & HC$_3$N (t) & $25-24$ & 141.9 & 227418.906 & 3.7 & -19.4 & 18.6 & 56.2$\pm$   9.9 & Tentative. \\
41 & $^{30}$SiS & $13-12$ & 76.5 & 227589.859 & 3.1 & -26.0 & 14.6 & 60.8$\pm$  12.8 &  \\
42 & CO & $2-1$ & 16.6 & 230538.000 & 0.4 & -37.6 & 24.2 & 58918.6$\pm$1654.7 &  \\
43 & $^{13}$CS & $5-4$ & 33.3 & 231220.688 & 3.1 & -20.0 & 21.6 & 127.1$\pm$  18.1 &  \\
44 & SiC$_2$ & $10_{2,9}-9_{2,8}$ & 69.6 & 232534.062 & 1.9 & -28.7 & 16.7 & 100.8$\pm$  32.6 &  \\
45 & SiC$_2$ & $10_{6,5}-9_{6,4}$ & 132.5 & 235713.000 & 4.8 & -22.9 & 23.4 & 56.6$\pm$ 17.8 & Doublet blend. \\
46 & SiC$_2$ & $10_{6,4}-9_{6,3}$ & 132.5 & 235713.062 & 4.8 & -22.8 & 23.5 & 56.6$\pm$  17.8 & Doublet blend. \\
47 & SiS & $13-12$ & 79.3 & 235961.359 & 1.5 & -33.0 & 31.5 & 832.0$\pm$ 112.2 & Irregular shape. \\
48 & HC$_3$N (t) & $26-25$ & 153.2 & 236512.781 & 2.3 & -22.6 & 24.7 & 41.2$\pm$  24.1 & Tentative. \\
49 & SiC$_2$ & $10_{4,7}-9_{4,6}$ & 93.7 & 237150.016 & 4.2 & -24.6 & 19.1 & 50.5$\pm$  18.2 &  \\
50 & SiC$_2$ & $10_{4,6}-9_{4,5}$ & 93.7 & 237331.312 & 4.2 & -20.3 & 21.2 & 90.4$\pm$  15.2 &  \\
51 & Im(SiO) & $/$ & -- & 239295.000 & -- & -- & -- & -- & Image contamination (Table~\ref{tbl:leakage}). \\
52 & CS & $5-4$ & 35.3 & 244935.562 & 0.8 & -33.6 & 19.9 & 3272.4$\pm$ 230.3 &  \\
53 & $^{30}$SiS (t) & $14-13$ & 88.2 & 245088.391 & 4.6 & -12.6 & 15.9 & 28.2$\pm$   5.3 & Tentative. \\
54 & HC$_3$N (t) & $27-26$ & 165.0 & 245606.312 & 2.4 & -15.7 & 24.4 & 63.6$\pm$  15.4 & Tentative. \\
55 & Im(HCN) & $/$ & -- & 246125.000 & -- & -- & -- & -- & Image contamination (Table~\ref{tbl:leakage}). \\
56 & Si$^{34}$S & $14-13$ & 89.0 & 247146.234 & 3.1 & -32.4 & 17.0 & 77.3$\pm$   9.3 &  \\
57 & NaCl (t) & $19-18$ & 118.7 & 247239.734 & 4.9 & -49.2 & 63.1 & 28.9$\pm$  25.5 & Tentative. \\
58 & SiC$_2$ & $10_{2,8}-9_{2,7}$ & 72.1 & 247529.125 & 4.2 & -18.9 & 18.0 &110.2$\pm$  15.7 &  \\
59 & u & $/$ & -- & 249401.000 & 2.4 & -58.9 & 20.3 & 132.8$\pm$  88.8 & No candidate. \\
60 & $^{29}$SiS & $14-13$ & 89.8 & 249435.406 & 2.4 & -15.4 & 14.7 & 65.7$\pm$   7.6 &  \\
61 & Si$^{17}$O & $6-5$ & 42.1 & 250744.688 & 4.9 & -13.4 & 23.1 & 58.9$\pm$   2.7 &  \\
62 & SiS & $14-13$ & 91.5 & 254103.219 & 2.3 & -31.7 & 18.0 & 887.1$\pm$ 190.1 &  \\
63 & $^{30}$SiO & $6-5$ & 42.7 & 254216.656 & 3.1 & -18.4 & 17.5 & 491.9$\pm$  61.0 &  \\
64 & SiC$_2$ (t) & $11_{2,10}-10_{2,9}$ & 81.9 & 254981.500 & 5.3 & -21.8 & 27.7 & 215.7$\pm$ 92.9 & Tentative. \\
65 & NaCl (t) & $20-19$ & 131.2 & 260223.109 & 4.9 & -46.8 & 65.8 & 5.3$\pm$   7.1 & Tentative. \\
66 & SiO & $6-5$ & 43.8 & 260518.016 & 1.5 & -28.1 & 28.7 & 4905.4$\pm$ 280.9 &  \\
67 & SiC$_2$ & $11_{4,8}-10_{4,7}$ & 106.2 & 261150.688 & 5.3 & -24.8 & 24.3 & 67.0$\pm$   10.3 &  \\
68 & HN$^{13}$C & $3-2$ & 25.1 & 261263.516 & 4.9 & -39.5 & 21.4 & 39.2$\pm$  12.6 & Tentative. \\
69 & SiC$_2$ & $11_{4,7}-10_{4,6}$ & 106.2 & 261509.328 & 3.8 & -23.0 & 21.7 & 59.7$\pm$  13.7 &  \\
70 & SiN & $N=6-5, J=11/2-9/2$ & 43.9 & 261650.000 & 1.5 & -21.5 & 14.6 &162.5$\pm$  35.6 &  \\
71 & SiC$_2$ & $12_{0,12}-11_{0,11}$ & 83.9 & 261990.750 & 2.3 & -16.3 & 18.5 & 200.5$\pm$29.7 & Blend with C$_2$H. \\
72 & C$_2$H & $N=3-2, J=7/2-5/2$ & 25.1 & 262005.000 & 2.3 & -27.9 & 37.2 & 490.4$\pm$ 47.2 & Hyperfine structure. Blend with SiC$_2$. \\
73 & C$_2$H & $N=3-2, J=5/2-3/2$ & 25.2 & 262067.000 & 2.3 & -54.0 & 21.3 & 294.7$\pm$  23.6 & Hyperfine structure. \\
74 & SiN & $N=6-5, J= 13/2-11/2$ & 44.1 & 262155.000 & 2.3 & -27.0 & 13.3 & 150.4$\pm$  27.0 &  \\
75 & C$_2$H (t) & $N=3-2, J=5/2-5/2$ & 25.2 & 262208.000 & 6.9 & -63.7 & 8.6 & 21.6$\pm$   10.9 & Hyperfine structure. Tentative. \\
76 & $^{30}$SiS & $15-14$ & 100.8 & 262585.031 & 3.1 & -30.9 & 12.1 & 117.1$\pm$  20.2 & Asymmetric. \\
77 & HC$_3$N (t) & $29-28$ & 189.9 & 263792.312 & 4.9 & -22.9 & 18.0 & 41.3$\pm$   8.3 & Tentative. \\
78 & Si$^{34}$S & $15-14$ & 101.7 & 264789.719 & 3.1 & -15.7 & 9.0 & 60.2$\pm$  13.2 &  \\
79 & HCN, $v_2=1$ & $3_{1}-2_{-1}$ & 1049.9 & 265852.719 & 1.5 & -12.4 & 19.3 & 272.5$\pm$  13.8 &  \\
80 & HCN & $3-2$ & 25.5 & 265886.438 & 1.5 & -33.8 & 25.7 & 12283.4$\pm$ 465.7 &  \\
81 & Im(CS) & $/$ & -- & 267070.000 & -- & -- & -- & -- & Image contamination (Table~\ref{tbl:leakage}). \\
82 & HCN, $v_2=1$ & $3_{-1}-2_{1}$ & 1050.0 & 267199.281 & 3.1 & -17.0 & 15.4 & 97.5$\pm$   9.9 &  \\
83 & $^{29}$SiS & $15-14$ & 102.6 & 267242.219 & 3.1 & -8.0 & 15.8 & 41.6$\pm$   4.6 &  \\
84 & HCO$^+$ (t) & $3-2$ & 25.7 & 267557.625 & 2.4 & -26.0 & 28.7 & 95.4$\pm$  29.0 & Tentative. \\
\hline
\end{longtable}

%% file: Mytblfoot.tex
\tblfootcitation {schoeier2013_hcn}
\tblfootcitation {decin2010_nlte}
\tblfootcitation {schoeier2011_chicyg}
\tblfootcitation {danilovich2014}
\tblfootcitation {agundez2012_innerlayers_irc10216}
\tblfootcitation {danilovich2018_cs_sis}
\tblfootcitation {brunner2018_waql_alma}
\tblfootcitation {massalkhi2019_cs_sio_sis}
\tblfootcitation {debeck2018_rdor}
\tblfootcitation {gonzalezdelgado2003}
\tblfootcitation {ramstedt2009_sio}
\tblfootcitation {schoeier2006_sio}
\tblfootcitation {maercker2016_water}
\tblfootcitation {wong2018_nh3}
\tblfootcitation {schmidt2016_irc10216_nh3}
\tblfootcitation {agundez2009_phd}
\tblfootcitation {massalkhi2018_sic2}
\tblfootcitation {debeck2012_cch}
\tblfootcitation {velillaprieto2017_iktau_iram}
\tblfootcitation {daniel2012_hnc}
\tblfootcitation {danilovich2017_h2s}
\tblfootcitation {debeck2013_popn}
\tblfootcitation {pulliam2011_hcoplus}
\tblfootcitation {danilovich2016_sulphur}